\documentclass[twocolumn]{pasj02}
\usepackage[switch,mathlines]{lineno} 
\usepackage{comment}
\usepackage{xcolor}
\usepackage{ulem}

\jyear{2025}
\Received{}
\Accepted{}

\graphicspath{{./}{figures/}} 

\begin{document} 

\title{A systematic study on the aromatic and aliphatic hydrocarbon emission features of nearby galaxies using AKARI near-IR spectra}

\author{
 Tsubasa \textsc{kondo},\altaffilmark{1}\altemailmark\orcid{0009-0002-5486-0068} \email{t.kondo@u.phys.nagoya-u.ac.jp} 
 Hidehiro \textsc{kaneda},\altaffilmark{1}
 Shinki \textsc{oyabu},\altaffilmark{2}
 Takuma \textsc{kokusho},\altaffilmark{1}
 Toyoaki \textsc{suzuki},\altaffilmark{3}
 Risako \textsc{katayama},\altaffilmark{1}
 Eiko \textsc{kozaki},\altaffilmark{1}
 Itsuka \textsc{yachi},\altaffilmark{1}
 Keita \textsc{yoshida},\altaffilmark{1}
 and
 Shohei \textsc{ono}\altaffilmark{1}
}
\altaffiltext{1}{Graduate School of Science, Nagoya University, Furo-cho, Chikusa-ku, Nagoya, Aichi 464-8602, Japan}
\altaffiltext{2}{Institute of Liberal Arts and Sciences, Tokushima University, 1-1 Minami-josanjima-cho,
Tokushima-shi, Tokushima, 770-8502, Japan}
\altaffiltext{3}{Institute of Space and Astronautical Science, Japan Aerospace Exploration Agency, 3-1-1 Yoshinodai, Chuo-ku, Sagamihara, Kanagawa 252-5210, Japan}




\maketitle

\begin{abstract}
Interstellar hydrocarbon dust containing aromatic and aliphatic hydrocarbons, like polycyclic aromatic hydrocarbons (PAHs), is believed to be processed by various factors including UV radiation fields and mechanical shocks in the galactic environments. We systematically investigate the processing of hydrocarbon dust, especially the likely causes for the variations of the luminosity ratio of aliphatic to aromatic hydrocarbon emission features, using the near-infrared (IR) spectral features at wavelengths 3.3\>\textmu m and 3.4--3.6\>\textmu m observed with AKARI/IRC. We analyzed 243 near-IR spectra of 240 star-forming (ultra-)luminous IR galaxies (the total IR luminosity, $L_\mathrm{IR}>10^{11}\ \mathrm{L_\odot}$), 119 spectra of 105 star-forming IR galaxies ($10^{10}\ \mathrm{L_\odot}<L_\mathrm{IR}<10^{11}\ \mathrm{L_\odot}$), and 94 spectra of 65 sub-IR galaxies ($L_\mathrm{IR}<10^{10}\ \mathrm{L_\odot}$), in addition to 232 spectra of 36 Galactic H\,\emissiontype{II} regions as a reference sample. We performed near-IR spectral model fitting to estimate the luminosities of the aromatic and aliphatic hydrocarbon features and the H\,\emissiontype{I} recombination line Br$\mathrm{\alpha}$. The result indicates that the luminosity ratios of the aliphatic to the aromatic hydrocarbons ($L_\mathrm{aliphatic}/L_\mathrm{aromatic}$) in the sample galaxies show considerably large variations, compared to those in the Galactic H\,\emissiontype{II} regions, $L_\mathrm{aliphatic}/L_\mathrm{aromatic}$ systematically decreasing with $L_\mathrm{IR}$ and $L_\mathrm{Br\alpha}$. We find that (sub-)IR galaxies with continuum colors bluer at 4~\textmu m tend to have higher $L_\mathrm{aliphatic}/L_\mathrm{aromatic}$, which is likely to reflect the intrinsic nature of PAHs outside the H\,\emissiontype{II} region where the PAHs remain non-processed by strong UV radiation fields. We also find that some ultra-luminous IR galaxies with continuum colors redder at 4~\textmu m show extremely low $L_\mathrm{aliphatic}/L_\mathrm{aromatic}$, which is likely to be caused by blending aliphatic emission and absorption features due to the presence of an obscured galactic nucleus in merger systems.
\end{abstract}


\section{Introduction}
Polycyclic aromatic hydrocarbons (PAHs) are believed to be likely carriers of the so-called unidentified infrared emission bands, whose spectral features are observed in a wide range of wavelengths in the near- to mid-infrared (IR) regions. Among them, representative ones are observed at wavelengths of 3.3, 6.2, 7.7, 8.6, 11.3, 12.7, and 17~\textmu m (e.g., \cite{d&l2007}; \cite{li2020}). For example, the aromatic hydrocarbon feature at 3.3~\textmu m is ascribed to a C-H stretching mode (e.g., \cite{draine2011}). At wavelengths of 3.4--3.6~\textmu m contiguous to the aromatic hydrocarbon feature at 3.3~\textmu m, there are sub-features produced by C-H stretching modes of aliphatic hydrocarbons (e.g., \cite{kwok2007}). Those features, especially the aromatic ones, have so far been detected from various kinds of objects including reflection nebulae, young stellar objects, Galactic H\,\emissiontype{II} regions, Galactic photo-dissociation regions, and external galaxies including not only star-forming but also quiescent galaxies (e.g., \cite{peeters2002}; \cite{kaneda2008}; \cite{yamagishi2012}; \cite{mori2014}). \\
\indent Several studies have reported that the processing of hydrocarbon dust is likely to be caused by various interstellar environmental factors such as radiation field intensity and radiation hardness (e.g., \cite{galliano2008}; \cite{tielens2008}; \cite{hemachandra2015}; \cite{lai2023}). For example, in M\>82, a nearby starburst galaxy, it is shown that the relative abundance of aliphatic hydrocarbons is higher in the halo region than in the galactic center region, suggesting that mechanical shocks by galactic superwinds may have shattered large carbonaceous grains to produce small aliphatic hydrocarbons (\cite{yamagishi2012}). On the other hand, in Galactic H\,\emissiontype{II} regions, the ratio of the aliphatic to the aromatic hydrocarbon intensity decreases with the PAH ionization degree which is affected by the UV radiation field. This is likely to reflect the fact that aliphatic hydrocarbons are more fragile than aromatic hydrocarbons for photodissociation (\cite{mori2014}). Similarly to Galactic H\,\emissiontype{II} regions, in luminous infrared galaxies (LIRGs) and ultra-luminous infrared galaxies (ULIRGs), the ratio of the aliphatic to the aromatic hydrocarbon luminosity decreases with the interstellar radiation field (ISRF) strength (\cite{kondo2024}). In addition to the ISRF strength, galaxy mergers with strong shocks may cause a further decrease in the ratio of the aliphatic to the aromatic hydrocarbons (\cite{kondo2024}). \\
\indent As for recent observations of JWST, in extragalactic H\,\emissiontype{II} regions of nearby star-forming galaxies (NGC\>628, NGC\>1365, NGC\>7496, and IC\>5332), \citet{egorov2023} show that the ratio of fluxes in JWST bands, $(F_\mathrm{F770W}+F_\mathrm{F1130W})/F_\mathrm{F2100W}$, tracing the mass fraction of PAHs to the total dust, decreases with the ionization parameter of the H\,\emissiontype{II} regions, suggesting that the extreme-UV radiation ($>$ 13.6~eV) destroys PAHs efficiently. In addition, the JWST observation of the nearby galaxy, NGC\>7469, reveals that the ratio of the aliphatic to the aromatic hydrocarbon intensity decreases with [Ne\,\emissiontype{III}]$/$[Ne\,\emissiontype{II}] that represents the hardness of the radiation field. This result indicates that the harder radiation fields destroy aliphatic hydrocarbons more efficiently than aromatic hydrocarbons (\cite{lai2023}). From a systematic study of galaxies at redshift $z\sim$ 0.2--0.5 observed with JWST NIRCam, \citet{lyu2025} show that the ratio of the aliphatic to the aromatic hydrocarbons decreases with the star formation rate (SFR) indicators ($L_\mathrm{IR}$, $L_\mathrm{aromatic}$, and spectral-energy-distribution-based SFR), but not with the redshift, stellar mass, metallicity, or morphology. \\
\indent Organic matter like PAHs is thought to evolve with changes in the molecular structure due to various kinds of processing in the interstellar environments (e.g., \cite{berne&tielens2012}). However, it is yet to be clarified how aromatic and aliphatic hydrocarbons are processed in various kinds of the interstellar environments of galaxies. In this study, systematically analyzing the near-IR spectra of the nearby galaxies observed by AKARI/IRC (Infrared Camera; \cite{murakami2007}; \cite{onaka2007}), we investigate how the luminosity ratio of the aliphatic to the aromatic hydrocarbons changes depending on the surrounding interstellar environments, and how the relationship between the ratio of the aliphatic to the aromatic hydrocarbons and the interstellar environments can be linked to the evolution of the organic matter.

\section{Observations}
\subsection{Spectroscopy with AKARI/IRC and data reduction}
The AKARI/IRC (\cite{onaka2007}) spectroscopic observations of nearby galaxies were carried out in the director's time program (DT), in the AKARI mission programs (MP) of AGNUL, AMUSE, EGANS, FUHYU, ISMGN, MSAGN, MSFGO, QSONG, and SPICY, and in the AKARI open time programs (OT) of ABCGA, ASCSG, BRSFR, CLNSL, COABS, GOALS, H2BCG, H2IRC, HOTCO, ISBEG, NISIG, NULIZ, SHARP, and SYDUS. The AKARI telescope and instruments were cooled by mechanical coolers and liquid helium (\cite{murakami2007}). We call the observation periods before and after the boil-off of the liquid helium Phase 2 and Phase 3, respectively. In our paper, we use the spectroscopic data of nearby galaxies retrieved from DARTS\footnote{https://darts.isas.jaxa.jp/en}. The spectral data were obtained by using the NG mode of the near-IR channel of IRC (2.5--5.0~\textmu m) in Phases 2 and 3 (\cite{ohyama2007}). We extracted spectral data from the slit (Ns or Nh) or the window (Np) as designated in the AOT parameter. \\
\indent For the data reduction,  we basically used the official AKARI data reduction pipeline for Phases 2 and 3 (IRC Spectroscopy Toolkit Version 20181203 and IRC Spectroscopy Toolkit for Phase 3 data Version 20181203)\footnote{https://www.ir.isas.jaxa.jp/AKARI/Observation/support/IRC/}. Since the temperature of the detector increased to $\sim$40~K in Phase 3, the number of hot pixels in the detector array increased, and accordingly the quality of the spectral data deteriorated. Therefore, in addition to the official AKARI data reduction pipeline, we applied custom procedures to all the spectral data taken in Phase 3 in order to remove the effect of the hot pixels and to improve the signal-to-noise ($S/N$) ratios (\cite{yamagishi2011}). Finally, we combined spectral data with the same target coordinates (RA, Dec) and the same AOT parameters. We selected nearby galaxies with redshift, $z<$ 0.4, since we focus on the 3.3~\textmu m aromatic hydrocarbon feature and the 3.4--3.6~\textmu m aliphatic hydrocarbon sub-features which are covered by AKARI/IRC near-IR spectroscopy. As a result, we newly analyzed 699 spectra of 623 nearby galaxies, in addition to the sample of \citet{kondo2024}, for the present study.

\subsection{Classification and selection of galaxies}
\label{data_cls_selection}
\begin{table*}[htbp]
\caption{Classification of the sample galaxies for the present study, based on the equivalent width of the 3.3~\textmu m aromatic hydrocarbon feature, $\mathrm{EW_{aromatic}}$, and the near-IR continuum slope, $\Gamma$.}
\label{list_sample_galaxy}
\centering
\begin{tabular}{lcccccc}
\hline
Sample & Star-forming & Composite & AGN & Others & \citet{kondo2024} & Total \\
\hline
ULIRG & 32 & N/A & N/A & N/A & 52 & 84 \\
LIRG & 85 & N/A & N/A & N/A & 71 & 156 \\
IRG & 94 & N/A & N/A & N/A & 11 & 105 \\
sub-IRG & 21 & 2 & 2 & 38 & 2 & 65 \\
\hline                 
\end{tabular}
\end{table*}

\begin{table*}[htbp]
\caption{Initial values and the range of the parameters in the model fitting.}
\label{list_parameters}
\centering
\begin{tabular}{ccccccc}
\hline
Feature & Central wavelength [\textmu m] & Feature width (Ns) & Feature width (Nh) & Feature width (Np) & Relative intensity \\ \hline
Aromatic & 3.29 (3.26--3.32) & 0.014 (0.012--0.035) & 0.014 (0.012--0.035) & 0.014 (0.012--0.035) & -- \\
Aliphatic 1 & 3.41 (3.39--3.44) & 0.030 (0.01--0.1)   & 0.030 (0.01--0.1)  & 0.030 (0.01--0.1) & -- \\
Aliphatic 2 & 3.46 (3.44--3.49) & 0.031 (0.025--0.050) & 0.025 (0.020--0.050) & 0.034 (0.030--0.050) & -- \\
Aliphatic 3 & 3.51 (3.49--3.53) & 0.031 (0.025--0.050) & 0.025 (0.020--0.050) & 0.034 (0.030--0.050) & -- \\
Aliphatic 4 & 3.56 (3.53--3.59) & 0.031 (0.025--0.050) & 0.025 (0.020--0.050) & 0.034 (0.030--0.050) & -- \\
\hline
H\emissiontype{I} Br$\mathrm{\beta}$ & 2.62 (2.59--2.65) & 0.031 (0.025--0.035) & 0.025 (0.020--0.030) & 0.034 (0.030--0.040) & 0.571 \\
H\emissiontype{I} Pf12 & 2.76 (2.73--2.79) & 0.031 (0.025--0.035) & 0.025 (0.020--0.030) & 0.034 (0.030--0.040) & 0.040 \\
H\emissiontype{I} Pf$\mathrm{\eta}$ & 2.88 (2.85--2.91) & 0.031 (0.025--0.035) & 0.025 (0.020--0.030) & 0.034 (0.030--0.040) & 0.050 \\
H\emissiontype{I} Pf$\mathrm{\epsilon}$ & 3.04 (3.01--3.07) & 0.031 (0.025--0.035) & 0.025 (0.020--0.030) & 0.034 (0.030--0.040) & 0.067 \\
H\emissiontype{I} Pf$\mathrm{\delta}$ & 3.30 & 0.031 (0.025--0.035) & 0.025 (0.020--0.030) & 0.034 (0.030--0.040) & 0.093 \\
H\emissiontype{I} Pf$\mathrm{\gamma}$ & 3.74 (3.71--3.77) & 0.031 (0.025--0.035) & 0.025 (0.020--0.030) & 0.034 (0.030--0.040) & 0.134 \\
$\mathrm{H_2}$ S(13) & 3.85 (3.82--3.88) & 0.031 (0.025--0.035) & 0.025 (0.020--0.030) & 0.034 (0.030--0.040) & -- \\
H\emissiontype{I} Hu15 & 3.91 (3.88--3.94) & 0.031 (0.025--0.035) & 0.025 (0.020--0.030) & 0.034 (0.030--0.040) & 0.012 \\
H\emissiontype{I} Br$\mathrm{\alpha}$ & 4.05 (4.02--4.08) & 0.031 (0.025--0.035) & 0.025 (0.020--0.030) & 0.034 (0.030--0.040) & 1.000 \\
H\emissiontype{I} Hu13 & 4.18 (4.15--4.21) & 0.031 (0.025--0.035) & 0.025 (0.020--0.030) & 0.034 (0.030--0.040) & 0.019 \\
He\emissiontype{I} & 4.30 (4.27--4.33) & 0.031 (0.025--0.035) & 0.025 (0.020--0.030) & 0.034 (0.030--0.040) & -- \\
H\emissiontype{I} Hu12 & 4.38 (4.35--4.41) & 0.031 (0.025--0.035) & 0.025 (0.020--0.030) & 0.034 (0.030--0.040) & 0.024 \\
H\emissiontype{I} Pf$\mathrm{\beta}$ & 4.65 (4.62--4.68) & 0.031 (0.025--0.035) & 0.025 (0.020--0.030) & 0.034 (0.030--0.040) & 0.202 \\
\hline
$\mathrm{H_2O}$ ice & 3.05 (2.90--3.10) & 0.13 (0.05--1.0) & 0.13 (0.05--1.0) & 0.13 (0.05--1.0) & -- \\
$\mathrm{CO_2}$ ice & 4.27 (4.22--4.32) & 0.15 (0.05--1.0) & 0.15 (0.05--1.0) & 0.15 (0.05--1.0) & -- \\
$\mathrm{CO}$ ice or gas & 4.67 (4.62--4.72) & 0.15 (0.05--1.0) & 0.15 (0.05--1.0) & 0.15 (0.05--1.0) & -- \\
\hline    
\end{tabular}
\end{table*}

\begin{figure}[htbp]
\centering
\includegraphics[bb=45 5 375 300, width=0.725\linewidth]{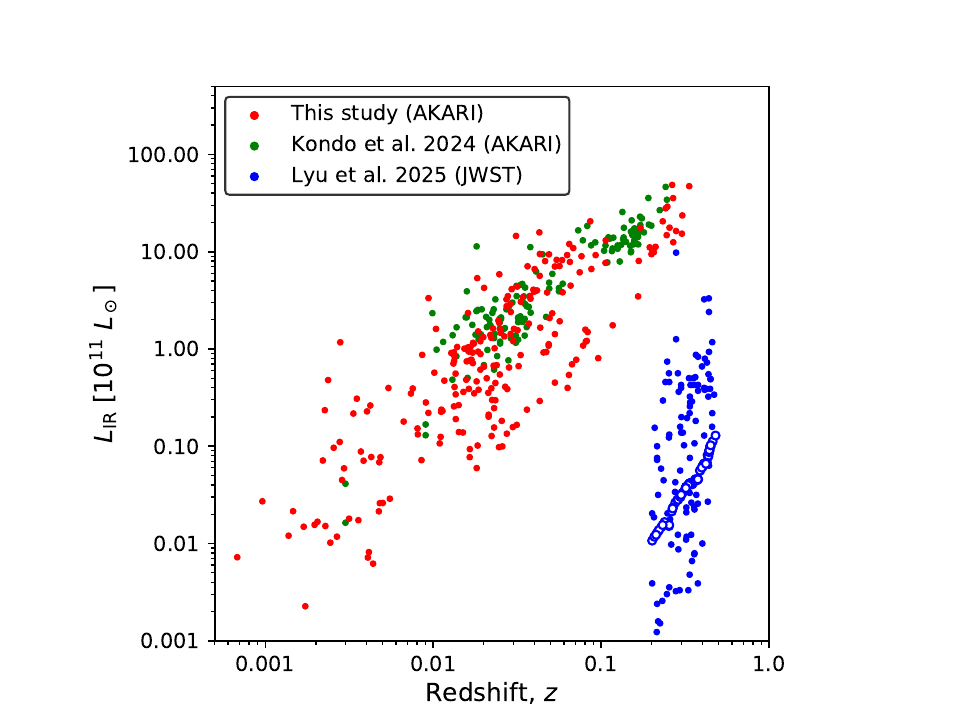}
\caption{Comparison of $L_\mathrm{IR}$ and redshift, $z$,  for the AKARI sample galaxies in this study (red circles), those in \citet{kondo2024} (green circles), and the JWST FRESCO sample galaxies in \citet{lyu2025} (blue circles), where open marks represent galaxies with $L_\mathrm{IR}$ upper limits. {Alt text: One scatter plot.}}
\label{lir_z_all_sample}
\end{figure}

We classify the sample galaxies into four luminosity classes based on the total infrared luminosity, $L_\mathrm{IR}$: (1) sub-infrared galaxies (sub-IRGs; $L_\mathrm{IR}<10^{10}\ \mathrm{L_\odot}$), (2) infrared galaxies (IRGs; $10^{10}\ \mathrm{L_\odot}<L_\mathrm{IR}<10^{11}\ \mathrm{L_\odot}$), (3) luminous infrared galaxies (LIRGs; $10^{11}\ \mathrm{L_\odot}<L_\mathrm{IR}<10^{12}\ \mathrm{L_\odot}$), and (4) ultra-luminous infrared galaxies (ULIRGs; $L_\mathrm{IR}>10^{12}\ \mathrm{L_\odot}$). $L_\mathrm{IR}$ is estimated from spectral energy distribution (SED) fitting, which is described in Section \ref{sed_fitting}. When it is difficult to estimate $L_\mathrm{IR}$ due to insufficient SED data points, we define the luminosity class of our sample galaxy referring to the information in the NASA/IPAC Extragalactic Database (NED)\footnote{https://ned.ipac.caltech.edu/} and also that in previous studies for several galaxies (\cite{s&m1996}; \cite{engelbracht2004}; \cite{kennicutt2011}; \cite{suzuki2013}; \cite{zhang2014}; \cite{persic2019}; \cite{persic2024}). \\
\indent We also classify the sample galaxies into three galaxy types based on the equivalent width of the 3.3~\textmu m aromatic hydrocarbon feature, $\mathrm{EW_{aromatic}}$, and the near-IR continuum slope, $\Gamma$ ($F_\mathrm{cont}\propto\lambda^\Gamma$ for the wavelength ranges of 2.55--2.70 and 3.60--4.00~\textmu m), as adopted in \citet{imanishi2010} and \citet{kondo2024}: (1) pure star-forming galaxies (SFGs; $\mathrm{EW_{aromatic}}>$ 40~nm and $\Gamma<$ 1), (2) SF-AGN (active galactic nucleus) composite galaxies $(\mathrm{EW_{aromatic}}>$ 40~nm and $\Gamma>$ 1), and (3) AGN-dominated galaxies ($\mathrm{EW_{aromatic}}<$ 40~nm and $\Gamma>$ 1). It should be noted, however, that this galaxy type classification method using near-IR spectra may misclassify, for example, compact obscured nuclei as pure SFGs because $\mathrm{EW_{aromatic}}$ of such AGNs can be as large as those of pure SFGs (e.g., \cite{garcia2025}). On the other hand, since the classification of sub-IRGs with $\mathrm{EW_{aromatic}}$ and $\Gamma$ is not straightforward because of the significant contribution of stellar components to the near-IR continuum in such IR-inactive galaxies, we treat all sub-IRGs as one group. As a result of this classification for the 623 nearby galaxies, we identify 216 pure star-forming (U)(L)IRGs and 103 sub-IRGs, on which we focus in this paper. AGN-inclusive (U)(L)IRGs are treated separately in \citet{katayama2025}. \\
\indent To estimate the fluxes of the hydrocarbon features, we carry out spectral model fitting, and then we select the spectra mainly based on the $\chi^2$ statistics for the full-range model fitting (section \ref{full_fit}) and /or the local-range (2.65--3.70~\textmu m) model fitting (section \ref{local_fit}) with the threshold of $\chi^2_\nu\leq$ 2.0. Since artifacts due to the data reduction and the instrumental errors tend to appear near the edges of the wavelength range of 2.5--5.0~\textmu m, we limit the fitting range to 2.55--4.85~\textmu m to avoid the influence of those artifacts. We find that the official AKARI data reduction pipeline produces unreasonably small or large spectral errors for the observed flux densities in some cases. Therefore, if necessary, we adjust error values by scaling them so that the error values are consistent with the standard deviation of the differential values between adjacent residual data points from the spectral fitting model in the wavelength ranges of 2.6--3.05 and 3.65--4.85~\textmu m where spectral features are relatively free, and then carry out spectral fitting again. \\
\indent For relatively faint spectra which show flux densities at 3.6--3.7~\textmu m less than 5.0~mJy and $\mathrm{EW_{aromatic}}$ less than 40~nm, where the detection of PAHs is susceptible to artifacts, we apply tight criteria using the result of the full-range model fitting (section \ref{full_fit}) to select the spectra that are relatively unaffected by artifacts. We also select galaxies with $\chi^2_\nu$ for the local-range model fitting less than 2.0. We add 30 spectra to the sample, which are rejected in the above data selection but significantly show strong hydrocarbon features without any apparent systematic residuals from the fitting model in the wavelength regions of 2.6--3.05 and 3.65--4.85~\textmu m. \\
\indent As a result, we obtain 228 spectra of 211 pure star-forming (U)(L)IRGs and 92 spectra of 63 sub-IRGs. Besides, we add the sample of \citet{kondo2024} which consists of 134 spectra of 134 pure star-forming (U)(L)IRGs and 2 spectra of 2 sub-IRGs. In addition, we re-analyze the 232 spectra of the 36 Galactic H\,\emissiontype{II} regions presented in \citet{mori2014} as a reference sample. The statistics of the sample galaxies are summarized in table \ref{list_sample_galaxy}. In figure \ref{lir_z_all_sample}, the distribution of $L_\mathrm{IR}$ as a function of redshift, $z$, for the sample galaxies in this study is compared with those of the AKARI sample in \citet{kondo2024} and the FRESCO sample observed with JWST/NIRCam WFSS (\cite{lyu2025}). The sample galaxies in this study span a wide range of $L_\mathrm{IR}$ at $z<$ 0.3, which are systematically fainter than those in \citet{kondo2024} and complementary to the FRESCO sample that covers much lower luminosities with a similarly broad $L_\mathrm{IR}$ range at $z>$ 0.2.

\section{Spectral fitting}
\label{spectral_fitting}
\subsection{Full-range model fitting}
\label{full_fit}
In order to exclude spectra with relatively large systematic errors in addition to evaluating errors in the spectra and correcting them if necessary, we first carry out a model fitting to the spectra in the wavelength range of 2.55--4.85~\textmu m, and then select the spectral data reliable for further analyses of hydrocarbon features (section \ref{local_fit}). The fitting model used here is given by

\begin{equation}
F_\nu=\mathrm{e}^{-(\tau_\mathrm{H_2O}+\tau_\mathrm{CO_2}+\tau_\mathrm{CO})}(F_\mathrm{cont}+F_\mathrm{aromatic}+F_\mathrm{aliphatic}+F_\mathrm{recomb}),
\label{all_model}
\end{equation}
where $\tau_\mathrm{H_2O}$, $\tau_\mathrm{CO_2}$, $\tau_\mathrm{CO}$, $F_\mathrm{cont}$, $F_\mathrm{aromatic}$, $F_\mathrm{aliphatic}$, and $F_\mathrm{recomb}$ are the optical depths due to $\mathrm{H_2O}$ ice, $\mathrm{CO_2}$ ice, and $\mathrm{CO}$ ice or gas, the flux densities of the continuum, the aromatic hydrocarbon feature at 3.3~\textmu m, the aliphatic hydrocarbon sub-features at 3.4--3.6~\textmu m, and the H\,\emissiontype{I} and He\,\emissiontype{I} recombination lines, respectively. For $F_\mathrm{cont}$, we use a polynomial function of the 4th or 5th order depending on the result of an $F$-test with a 90\% confidence level. For $F_\mathrm{aromatic}$, we use the following Drude profile (\cite{d&l2007}):
\begin{equation}
F_\mathrm{aromatic}=\frac{a_rb_r^2}{\left(\frac{\lambda}{\lambda_r}-\frac{\lambda_r}{\lambda}\right)^2+b_r^2},
\end{equation}
where $a_r$ and $b_r$ are the peak of the feature profile and the FWHM divided by $\lambda_r$, the central wavelength of the feature profile, respectively. \\
\indent The aliphatic hydrocarbon features at 3.4--3.6~\textmu m are considered to be composed of several components. Here we assume four representative components (e.g., \cite{sloan1997}, \cite{kwok2011}), while \citet{mori2014} assumed two components. Following \citet{kondo2024}, we apply a Lorentzian profile for the 3.41~\textmu m feature and Gaussian profiles for the 3.46, 3.51, and 3.56~\textmu m features. Since the H\,\emissiontype{I} and He\,\emissiontype{I} recombination emission lines are not spectrally resolved with the Ns or Nh slits or the Np window, we use Gaussian profiles to fit the emission lines. For the wavelength range of 2.55--4.85~\textmu m, we consider the 13 recombination lines listed in table \ref{list_parameters}:
\begin{equation}
F_\mathrm{recomb}=\sum^{13}_{i=1}\alpha_i\exp{\left[-4\ln{2\frac{(\lambda-\lambda_{r,\ i})^2}{\gamma^2}}\right]},
\end{equation}
where $\alpha$, $\gamma$, and $\lambda_r$ are the Gaussian profile peak, the FWHM, and the central wavelength, respectively. Finally, we use Gaussian profiles to fit the absorption features due to $\mathrm{H_2O}$ ice, $\mathrm{CO_2}$ ice, and $\mathrm{CO}$ ice or gas, following \citet{shimonishi2010}:
\begin{equation}
\tau_\mathrm{H_2O}+\tau_\mathrm{CO_2}+\tau_\mathrm{CO}=\sum^{3}_{i=1}\alpha_i\exp{\left[-4\ln{2\frac{(\lambda-\lambda_{r,\ i})^2}{\gamma^2}}\right]}.
\end{equation}

\begin{figure}[htbp]
\centering
\includegraphics[bb=30 15 525 380, width=0.75\linewidth]{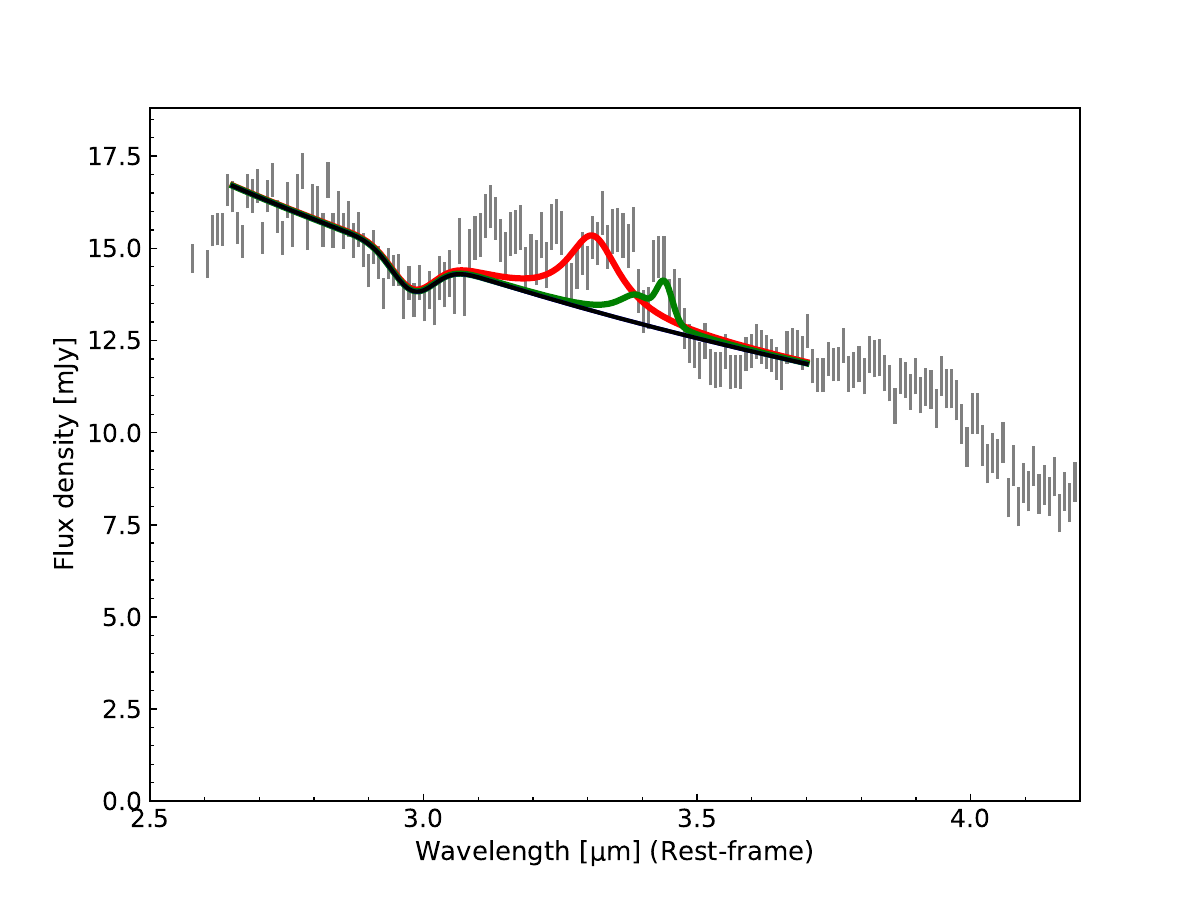}
\caption{Example spectrum showing an unusually broad feature over the wavelength range of 3.1--3.2~\textmu m. The black, red, and green lines represent the near-IR continuum, the 3.3~\textmu m aromatic feature, and the 3.4--3.6~\textmu m aliphatic feature, respectively. {Alt text: One example spectrum observed with AKARI/IRC.}}
\label{example_broad_feature}
\end{figure}

\indent In fitting the spectral data with the model, we divide the fitting procedure into three steps and determine the fitting parameters step by step. First, we carry out model fitting to determine the parameters of the continuum for the wavelength ranges of 2.55--2.60, 2.65--2.70, 3.60--3.70, 3.80--4.00, 4.10--4.20, 4.35--4.45, and 4.80--4.85~\textmu m, which are free of the hydrocarbon features, the H\,\emissiontype{I} and He\,\emissiontype{I} recombination lines, and the ice absorption features. Secondly, we determine parameters for the hydrocarbon features, the H\,\emissiontype{I} and He\,\emissiontype{I} recombination lines, and the ice absorption features, while the parameters of the continuum determined in the first fitting step are fixed. Finally, we fine-tune the amplitudes of the absorption features and all the emission features and lines under the condition that the continuum parameters and the shape parameters of the features and the lines (i.e., the feature and line widths, the central wavelength) are fixed at the values determined in the first and second fitting steps. We also fix the relative strengths of the four aliphatic components at those determined in the second step (i.e., the overall profile of the aliphatic feature. Here we confirm that the aliphatic flux error thus estimated is comparable to the sum of the observed flux density errors for the 3.4--3.6~\textmu m excess above the underlying continuum (the former-to-latter ratio of $1.3^{+0.9}_{-0.1}$ on average), and thus reasonable or at least not underestimated. We adopt such initial values and limited fitting ranges of the parameters as listed in table \ref{list_parameters}. The ranges of the amplitudes of the H\,\emissiontype{I} recombination lines are constrained within a factor of 2 with respect to the values predicted in the Case B condition ($T=10^4$~K, $n_\mathrm{e}=10^3\ \mathrm{cm^{-3}}$) on the basis of the Br$\mathrm{\alpha}$ line strength (\cite{h&s1987}, see table \ref{list_parameters}) except the Pf$\mathrm{\delta}$ line whose amplitude is fixed at the value predicted in the Case B condition, since the amplitude of the Pf$\mathrm{\delta}$ line cannot be determined due to blending with the aromatic hydrocarbon feature at 3.3~\textmu m. \\
\indent Based on the result of the model fitting, we evaluate a $\chi^2$ divided by the number of spectral data points ($\overline{\chi^2}$) over the wavelength ranges of 2.6--3.05 and 3.65--4.85~\textmu m, and then select the spectral data with the threshold of $\overline{\chi^2}\leq$ 2.0 to exclude spectra with relatively large systematic deviations from the fitting model due to artifacts. In some cases, spectral data show an unusually broad feature over the wavelength range of 3.05--3.2~\textmu m, which is likely a fake feature, but not a real hydrocarbon feature; we exclude the spectra that show the ratio of the integrated flux over the wavelength range of 3.05--3.2~\textmu m to the aromatic hydrocarbon feature at 3.3~\textmu m higher than 1$/$3. An example is shown in figure \ref{example_broad_feature}, where a broad excess is seen at 3.1--3.2~\textmu m with an FWHM of approximately 0.1~\textmu m. The only known feature or line around this wavelength range is H$\mathrm{_2}$ 1--0 O(5) at 3.235~\textmu m. However, since this excess is significantly broader than the spectral resolution of AKARI/IRC, it is likely to be a fake feature. Through this spectral fitting, we calculate the luminosities of the aromatic hydrocarbon emission feature ($L_\mathrm{aromatic}$), aliphatic hydrocarbon emission features ($L_\mathrm{aliphatic}$), and the H\,\emissiontype{I} recombination line Br$\mathrm{\alpha}$ ($L_\mathrm{Br\alpha}$). \\

\begin{figure*}[htbp]
\centering
\begin{tabular}{rcl}
\centering
\includegraphics[bb=35 15 525 400, clip, width=0.26\textwidth]{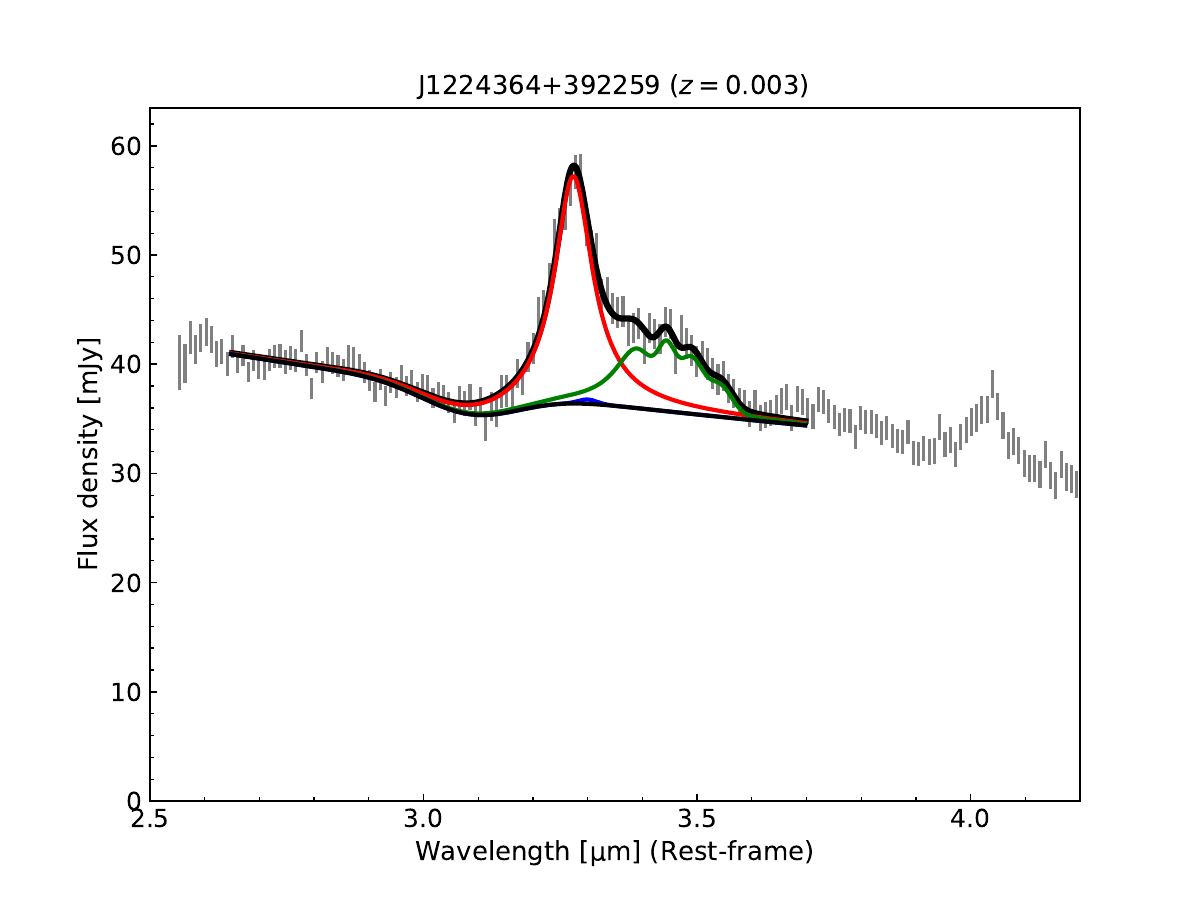} &
\includegraphics[bb=35 15 525 400, clip, width=0.26\textwidth]{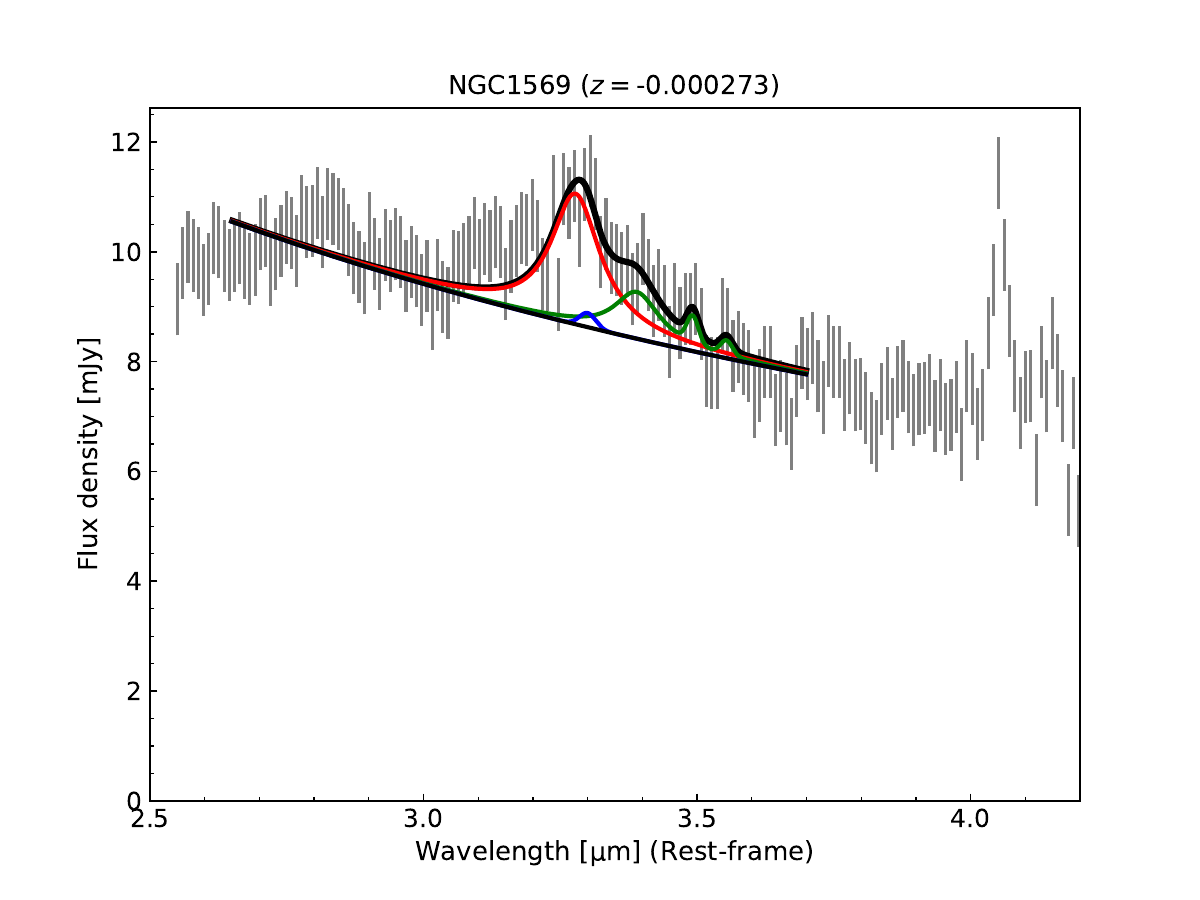} &
\includegraphics[bb=35 15 525 400, clip, width=0.26\textwidth]{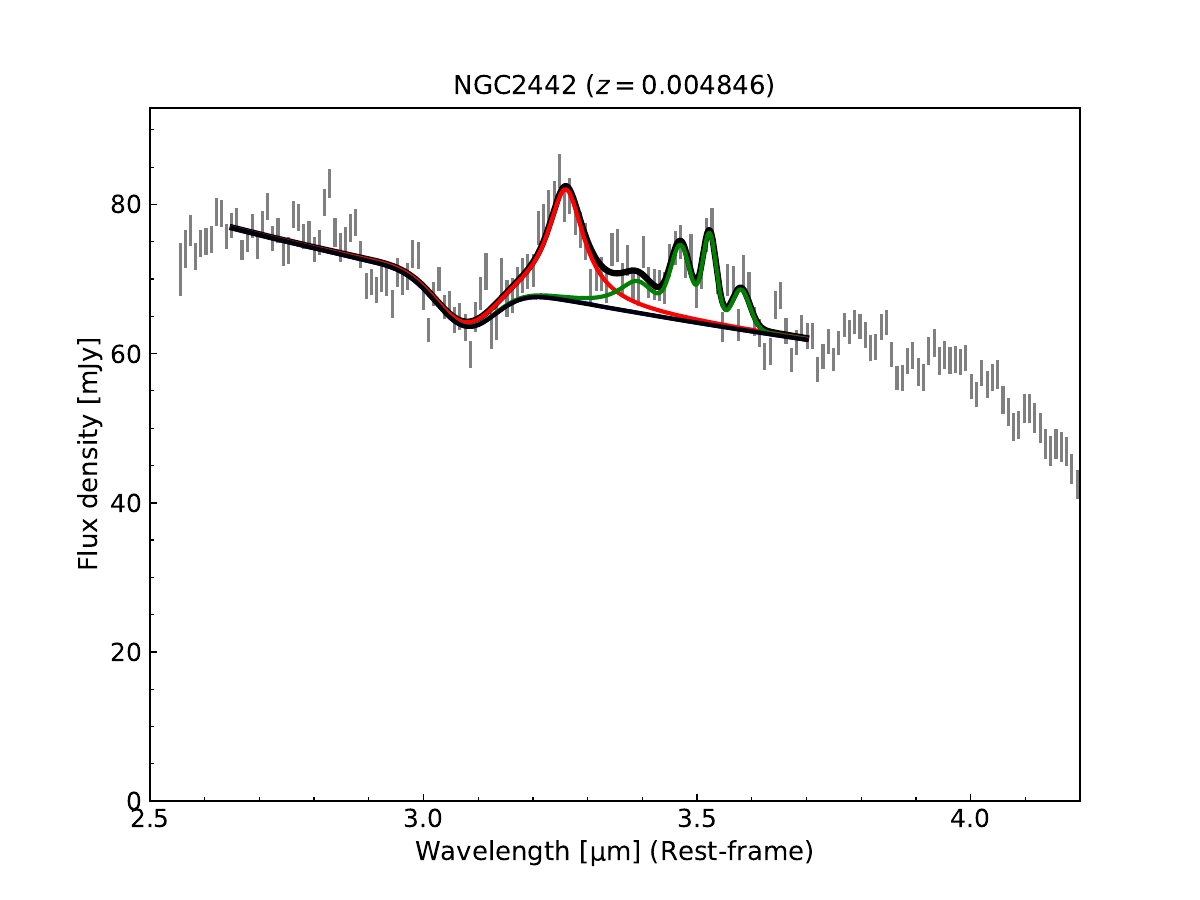} \\
\includegraphics[bb=35 15 525 400, clip, width=0.26\textwidth]{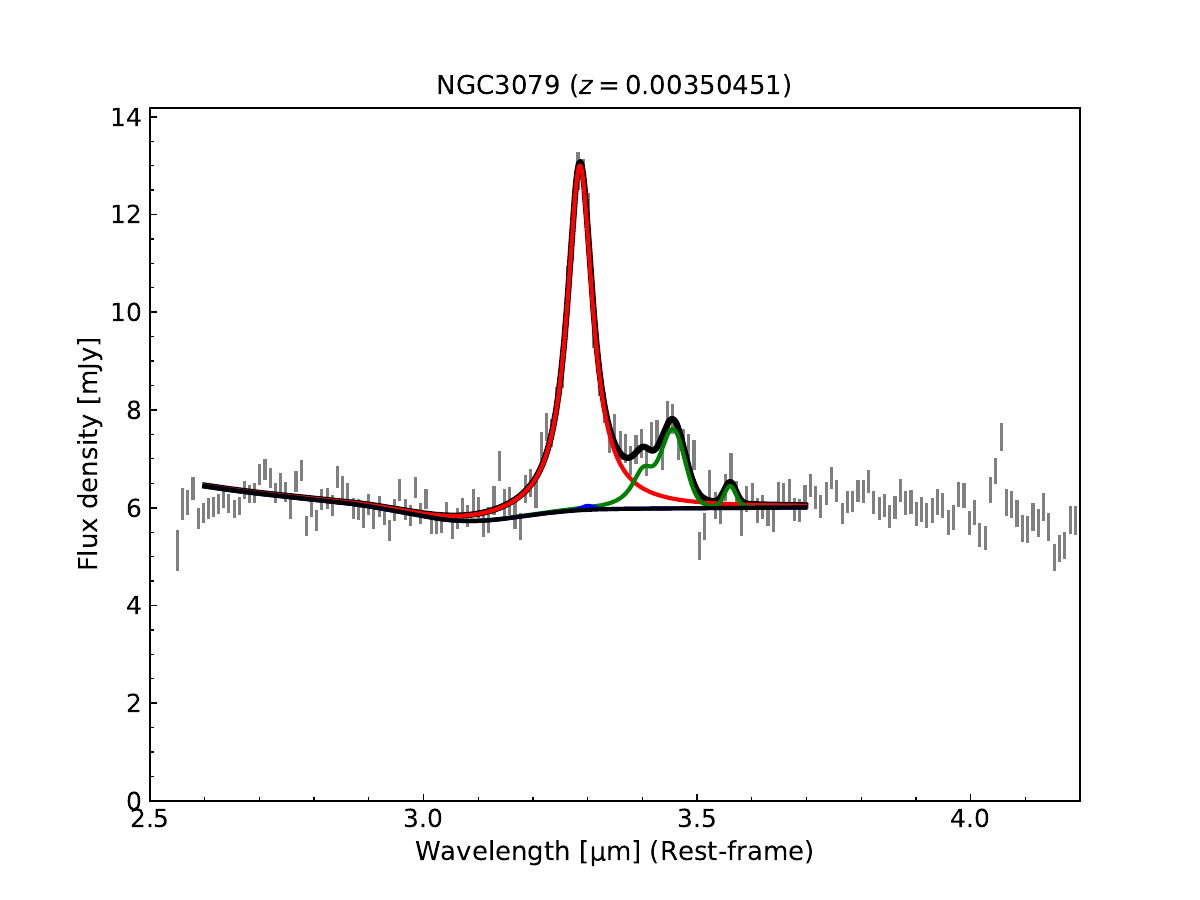} &
\includegraphics[bb=35 15 525 400, clip, width=0.26\textwidth]{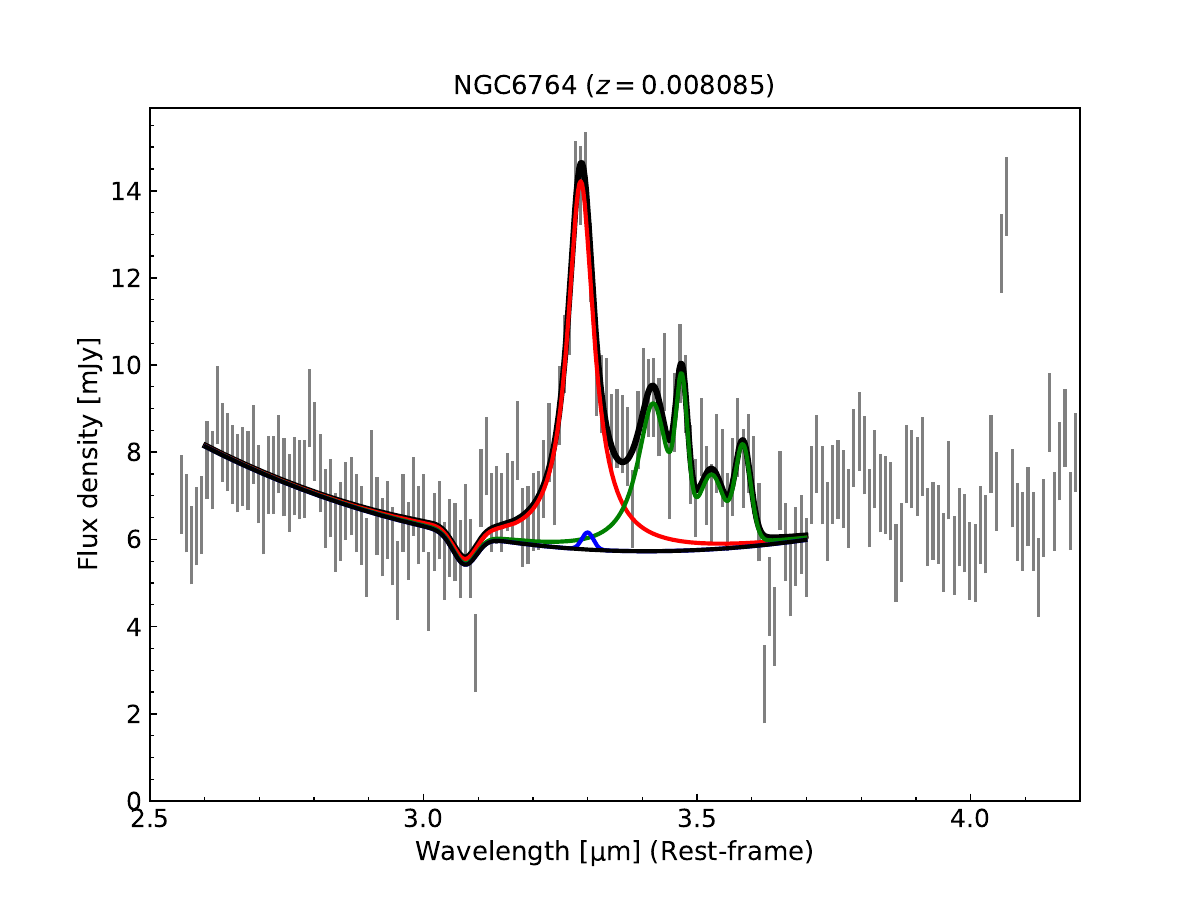} &
\includegraphics[bb=35 15 525 400, clip, width=0.26\textwidth]{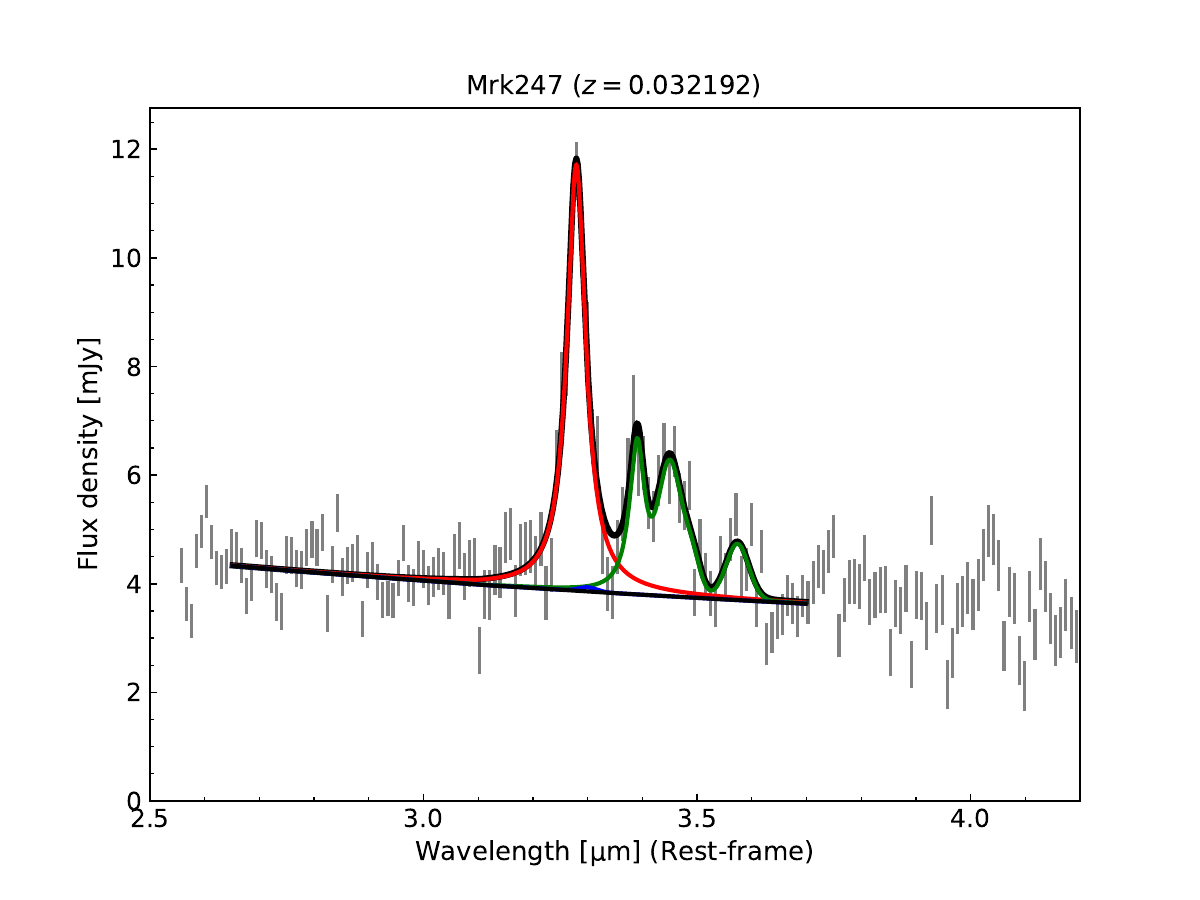} \\
\includegraphics[bb=35 15 525 400, clip, width=0.26\textwidth]{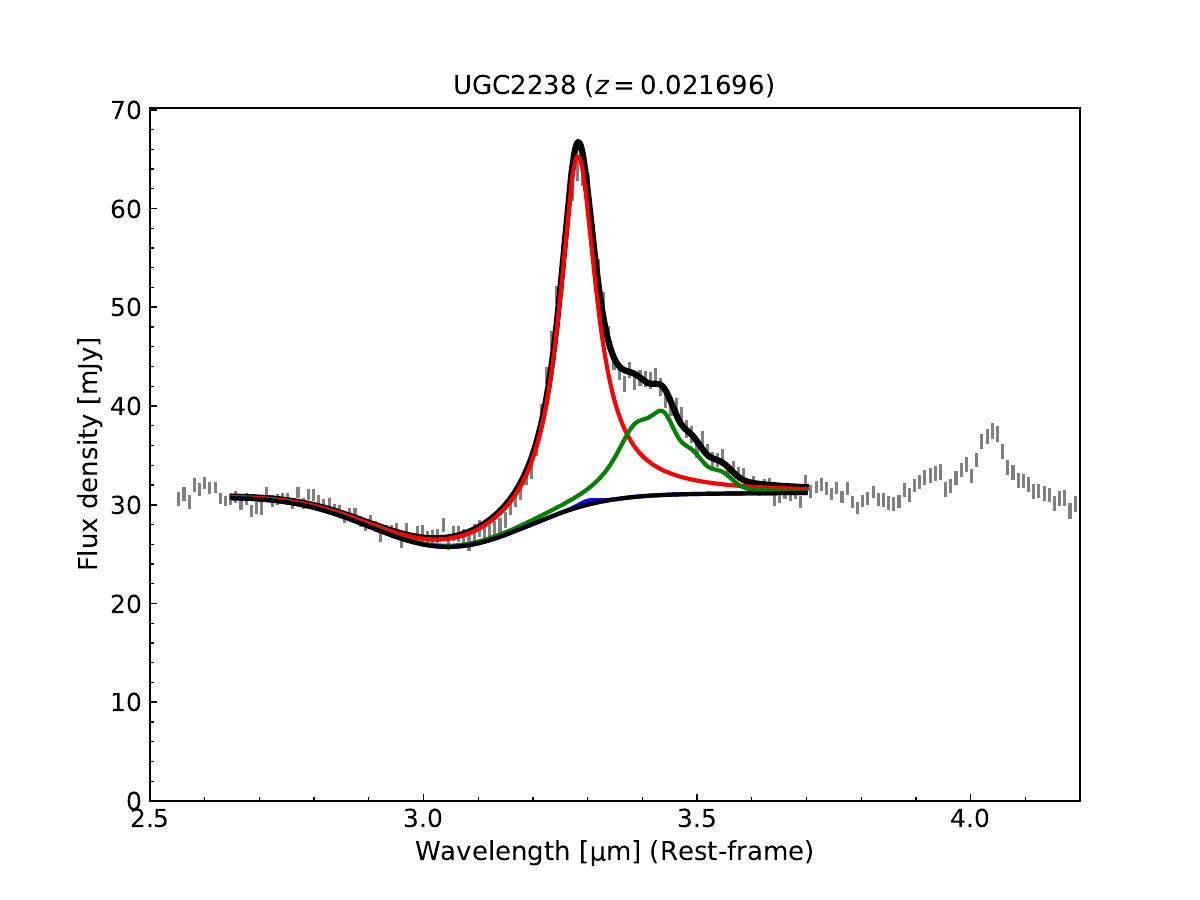} &
\includegraphics[bb=35 15 525 400, clip, width=0.26\textwidth]{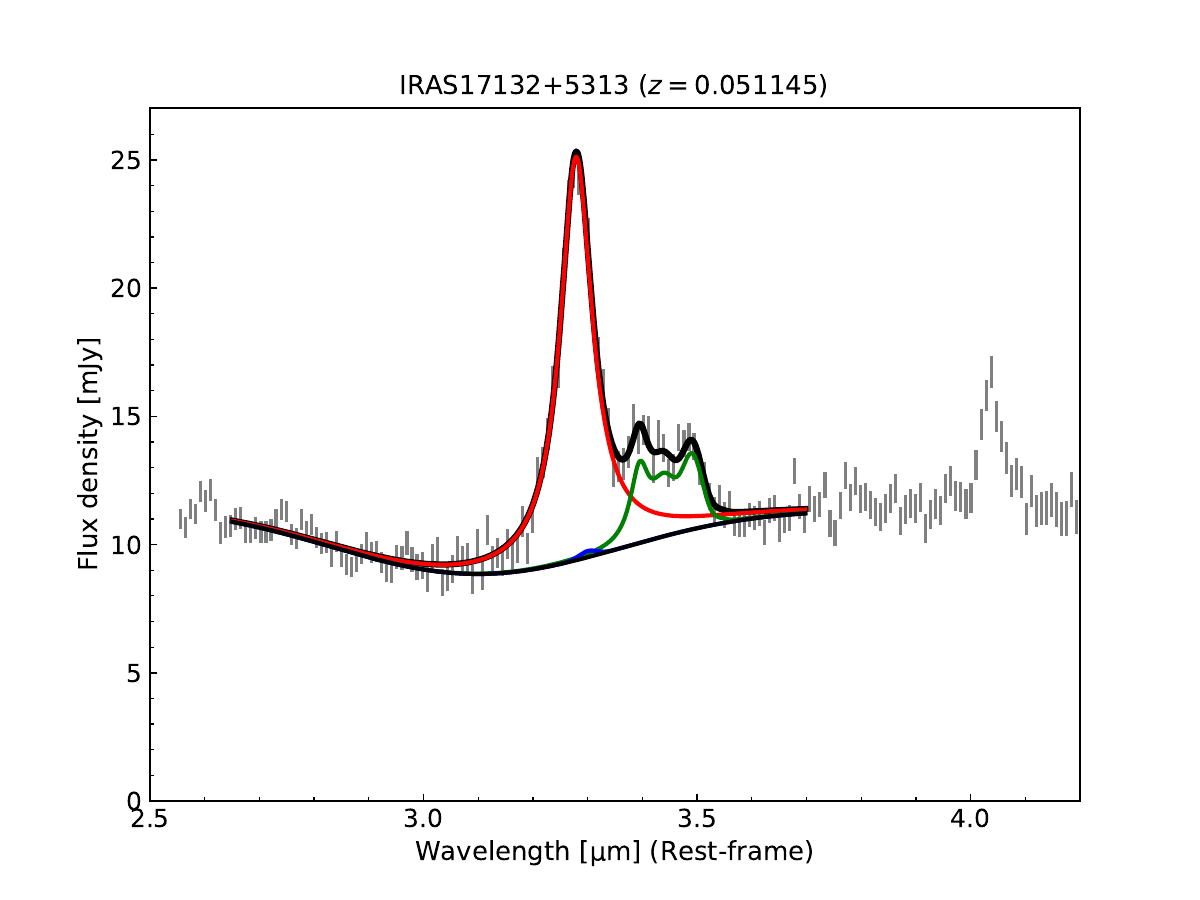} &
\includegraphics[bb=35 15 525 400, clip, width=0.26\textwidth]{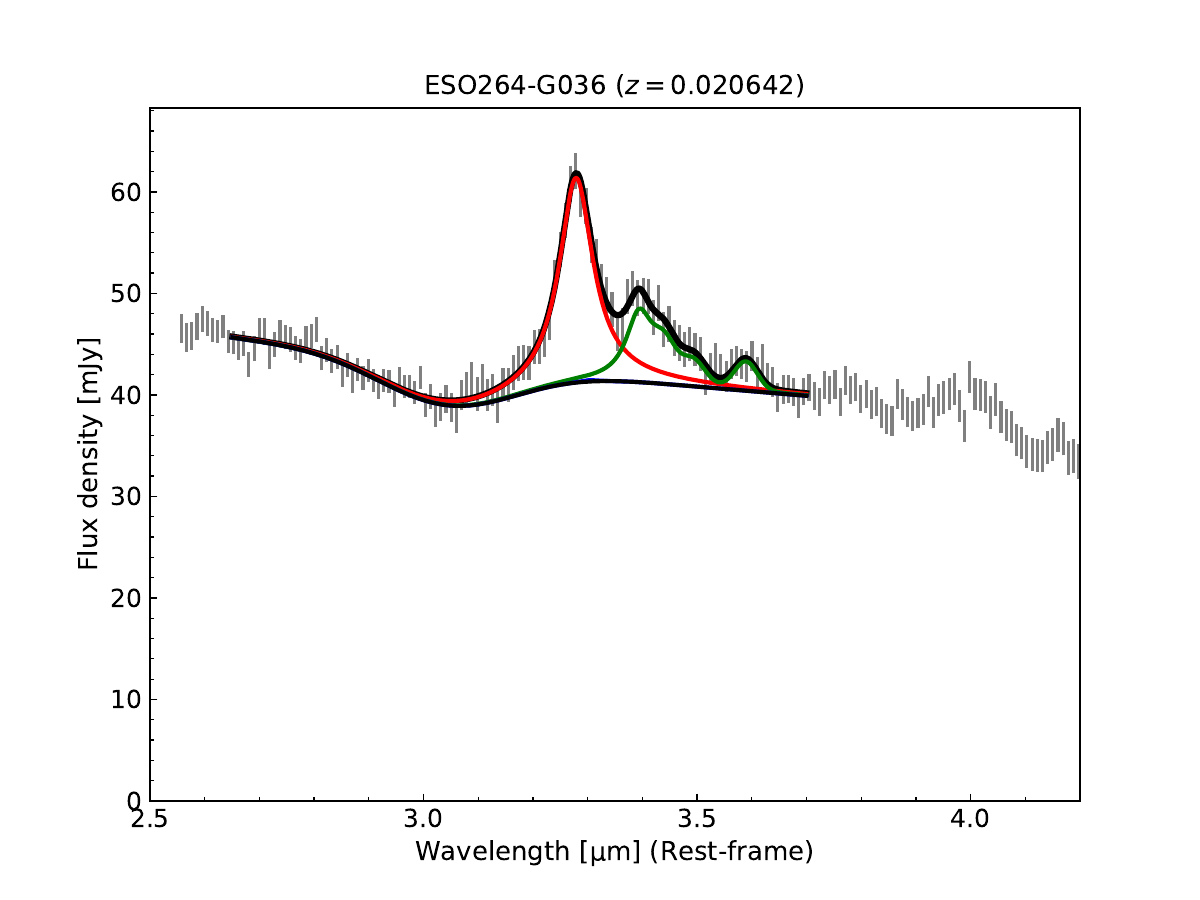} \\
\includegraphics[bb=35 15 525 400, clip, width=0.26\textwidth]{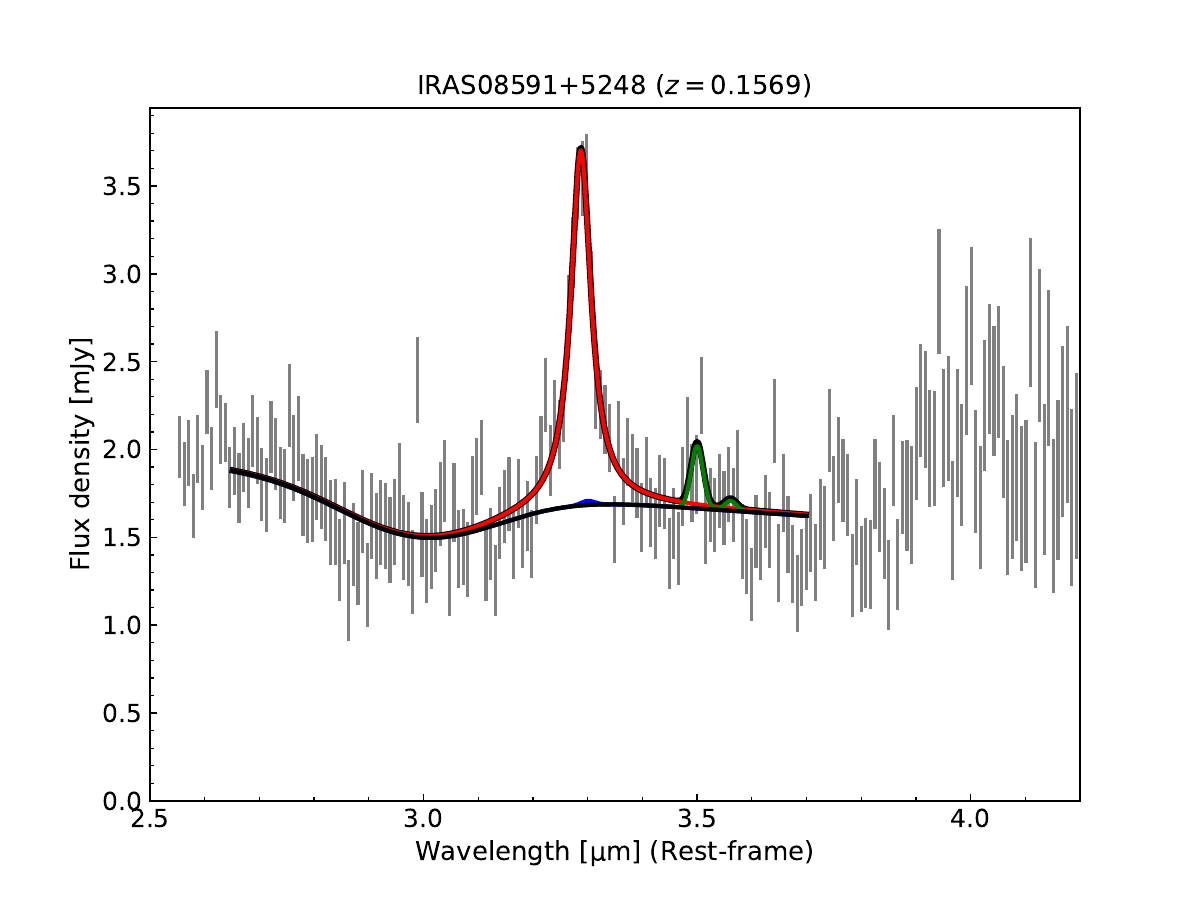} &
\includegraphics[bb=35 15 525 400, clip, width=0.26\textwidth]{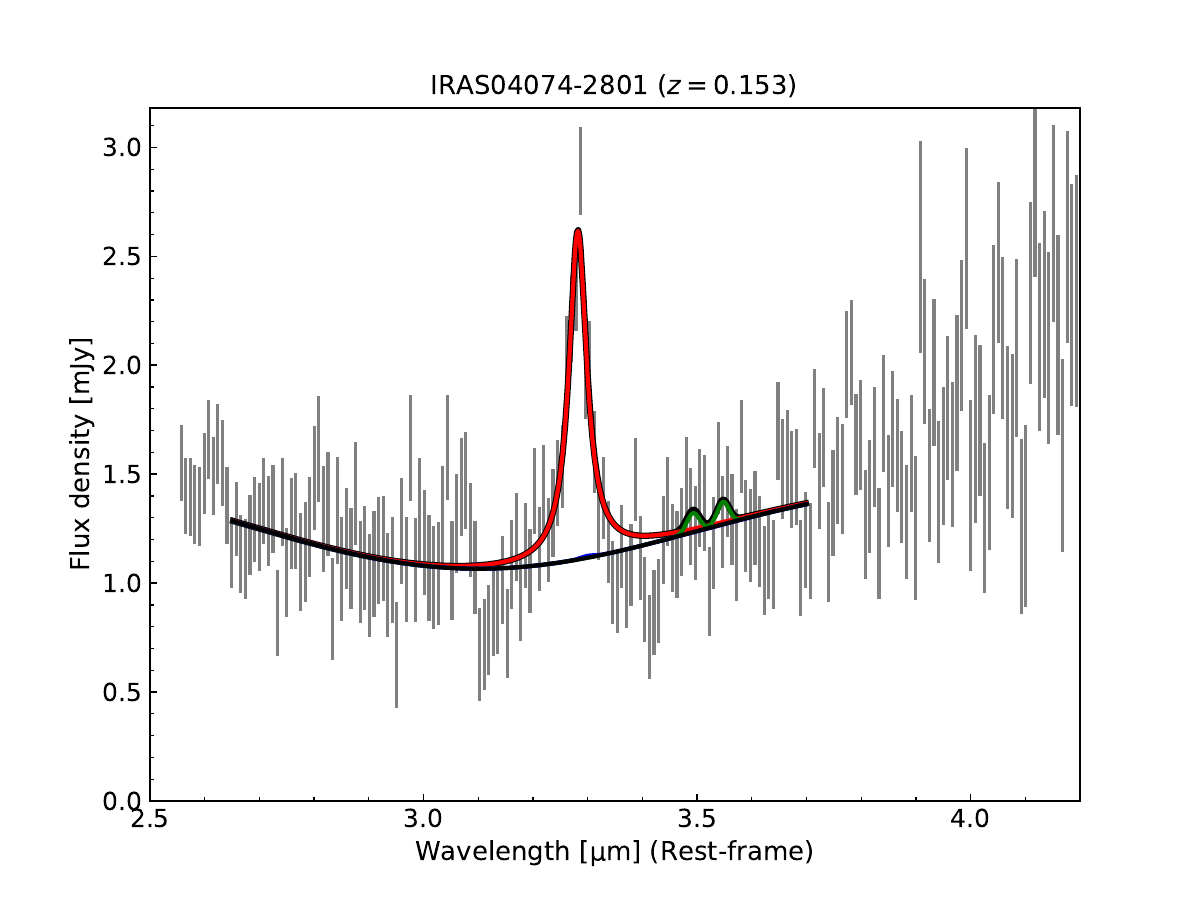} &
\includegraphics[bb=35 15 525 400, clip, width=0.26\textwidth]{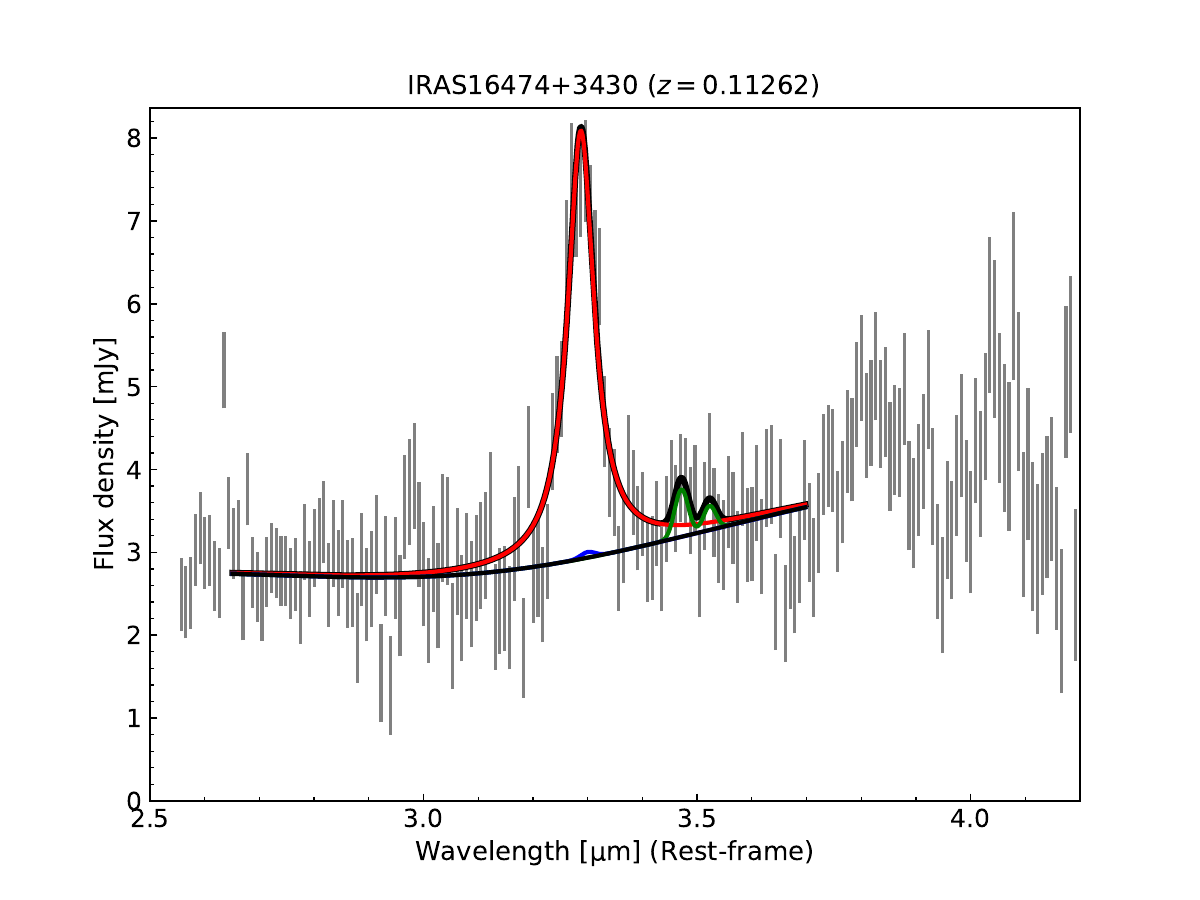}
\end{tabular}
\caption{Examples of the near-IR spectral fitting using the local-range model. The spectral model is composed of the near-IR continuum (black line), aromatic hydrocarbon emission feature (red line), aliphatic hydrocarbon emission features (green lines), and H\,\emissiontype{I} Pf$\mathrm{\delta}$ recombination line (blue line). The hydrocarbon features and Pf$\mathrm{\delta}$ recombination line are attenuated by the $\mathrm{H_2O}$ ice absorption. {Alt text: Twelve panels of examples for near-infrared spectral fitting.}}
\label{example_nir_fitting}
\end{figure*}

\subsection{Local-range model fitting}
\label{local_fit}
The full-range fitting model may have uncertainties in determining the shape of the underlying continuum below the aromatic and aliphatic features, because it is difficult to uniquely determine the full-range continuum with a single model. Those features are quite broad, while the Br$\mathrm{\alpha}$ line is narrow but its flux estimation is somewhat affected by the presence of the CO$_2$ absorption feature. Therefore, for the purpose of estimating the fluxes of the aromatic and aliphatic features more robustly, we perform local-range model fitting for the spectra with the corrected errors. Following \citet{kondo2024}, we use a power-law ($F_\mathrm{cont}\propto\lambda^\Gamma$, where $\Gamma$ is a free parameter) as a continuum model to determine the local-range continuum for the hydrocarbon features in the wavelength ranges of 2.55--2.60, 2.65--2.70, 3.60-3.70, and 3.80--4.00~\textmu m. On the other hand, since it is difficult to reproduce the local-range continuum using a power-law model when the hot dust emission is relatively strong, we also test the following model:
\begin{equation}
F_\mathrm{cont}=C_1\lambda^\Gamma+C_2\frac{B_\nu(\lambda, T)}{\lambda^2},
\end{equation}
where $C_1$, $C_2$ are the amplitudes of the power-law and the modified blackbody model, respectively. Based on the result of an $F$-test with a 90\% confidence level, we adopt the power-law or the second model as a continuum model for the local-range fitting. The fitting models for the aromatic hydrocarbon feature, the aliphatic hydrocarbon features, the H\,\emissiontype{I} recombination line Pf$\mathrm{\delta}$, and the $\mathrm{H_2O}$ ice absorption feature are the same as those used in the full-range fitting model, while the results of the full-range fitting are used as initial input values. Figure \ref{example_nir_fitting} shows examples of the result for the near-IR spectral model fitting with the local-range model. Through this spectral fitting, we calculate $L_\mathrm{aromatic}$ and $L_\mathrm{aliphatic}$, which are adopted in the following results and discussions.

\section{Spectral energy distribution fitting}
\label{sed_fitting}

\begin{figure*}[htbp]
\centering
\begin{tabular}{rcl}
\centering
\includegraphics[bb=5 0 430 325, clip, width=0.25\textwidth]{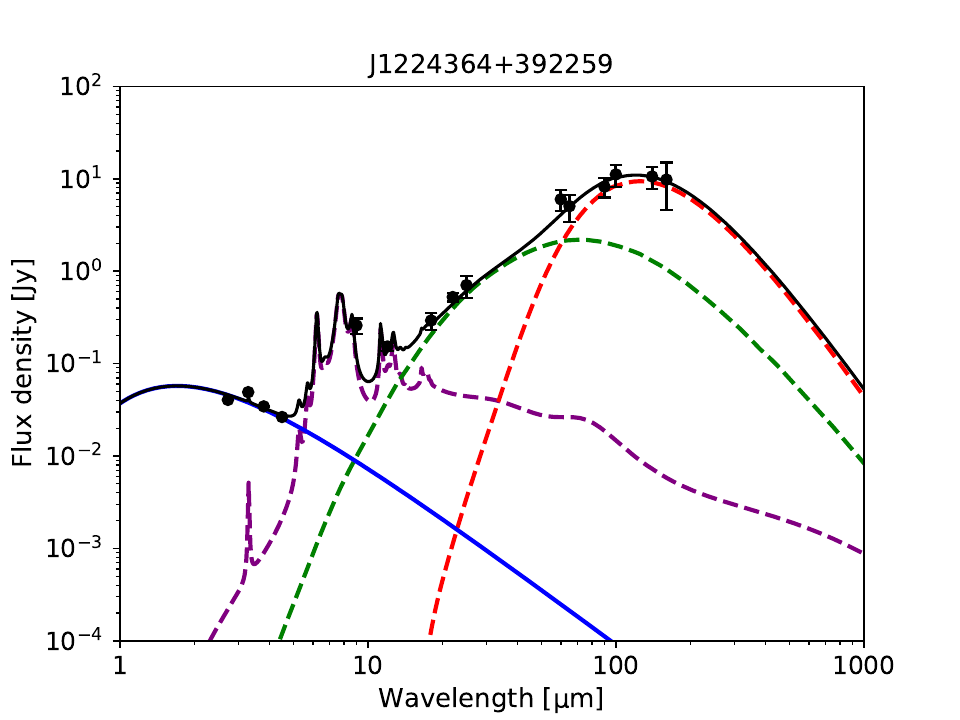} &
\includegraphics[bb=5 0 430 325, clip, width=0.25\textwidth]{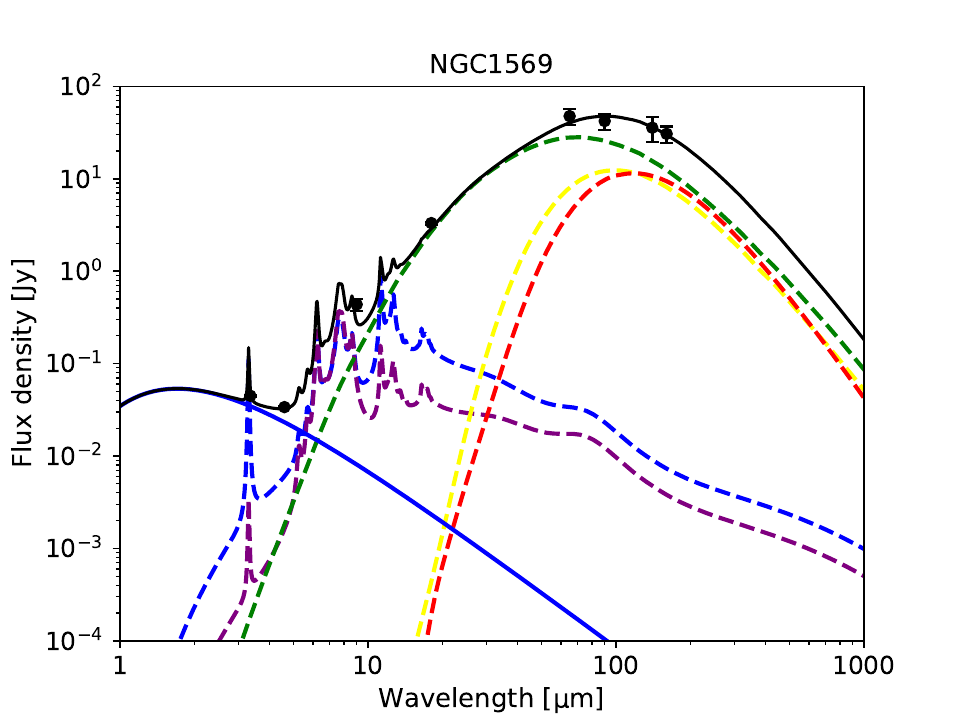} &
\includegraphics[bb=5 0 430 325, clip, width=0.25\textwidth]{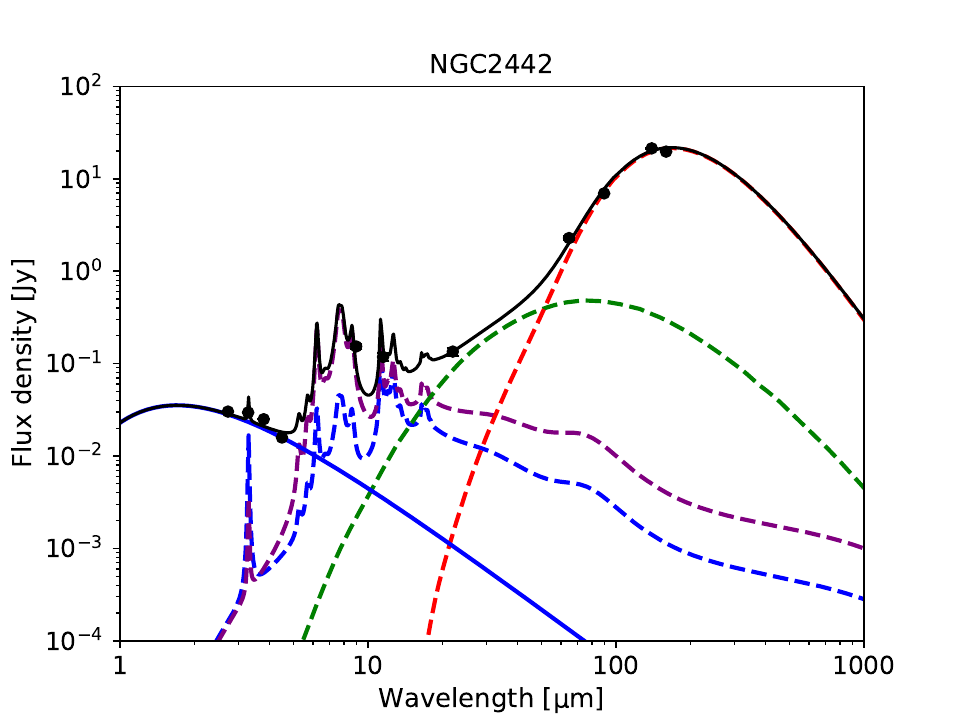} \\
\includegraphics[bb=5 0 430 325, clip, width=0.25\textwidth]{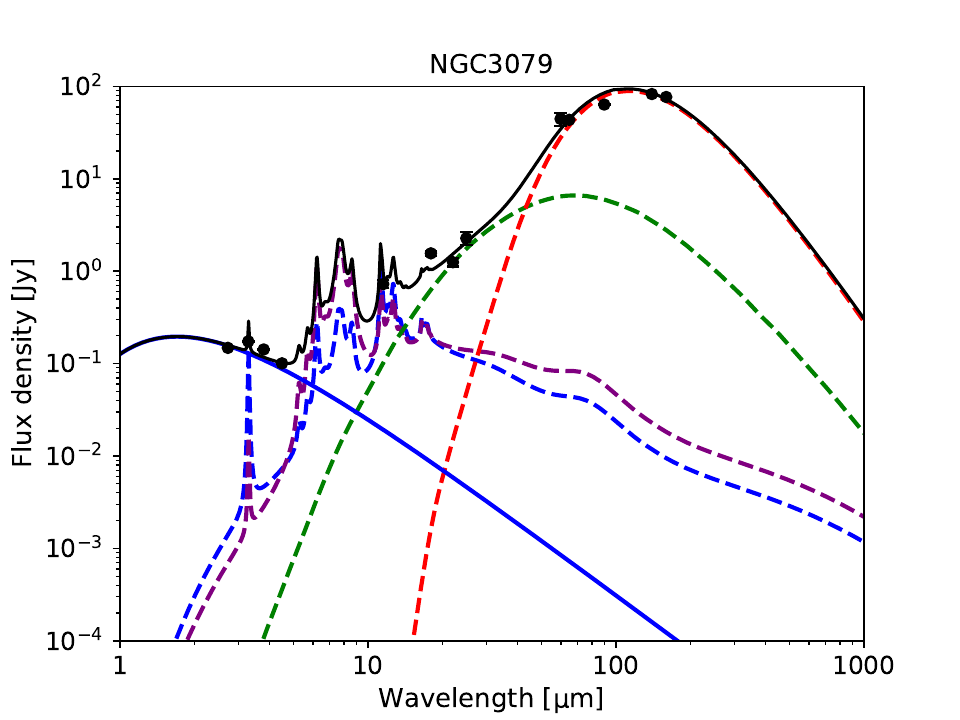} &
\includegraphics[bb=5 0 430 325, clip, width=0.25\textwidth]{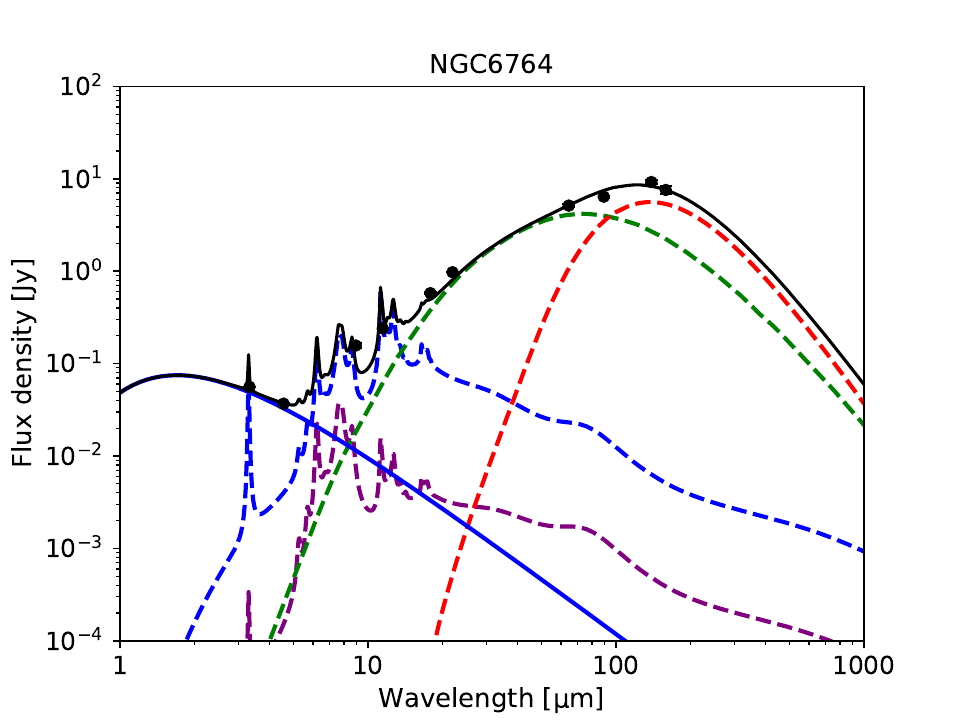} &
\includegraphics[bb=5 0 430 325, clip, width=0.25\textwidth]{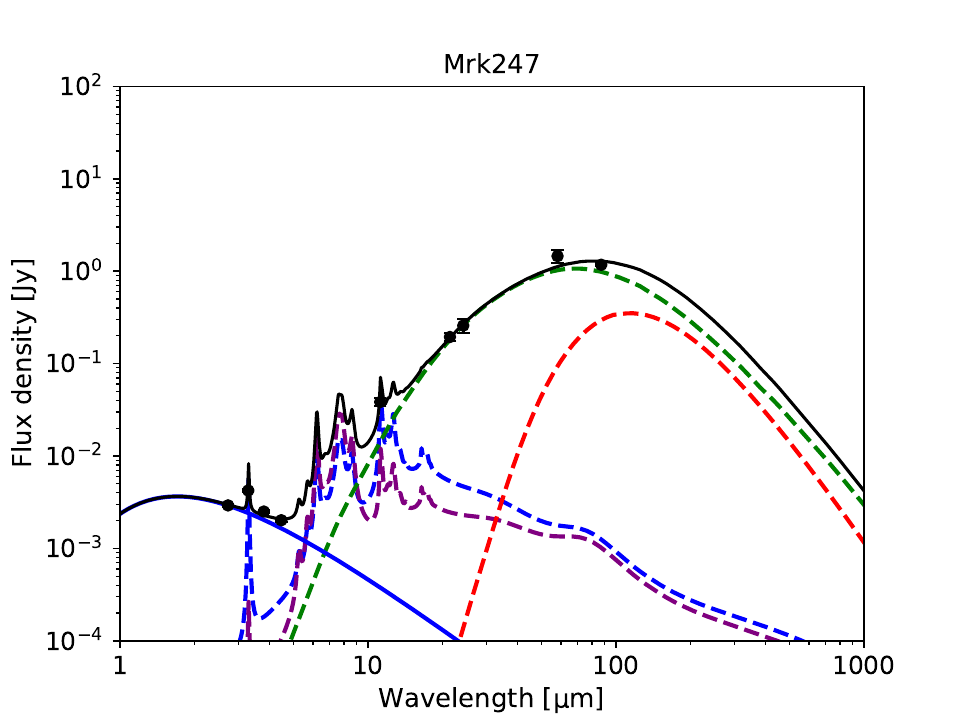} \\
\includegraphics[bb=5 0 430 325, clip, width=0.25\textwidth]{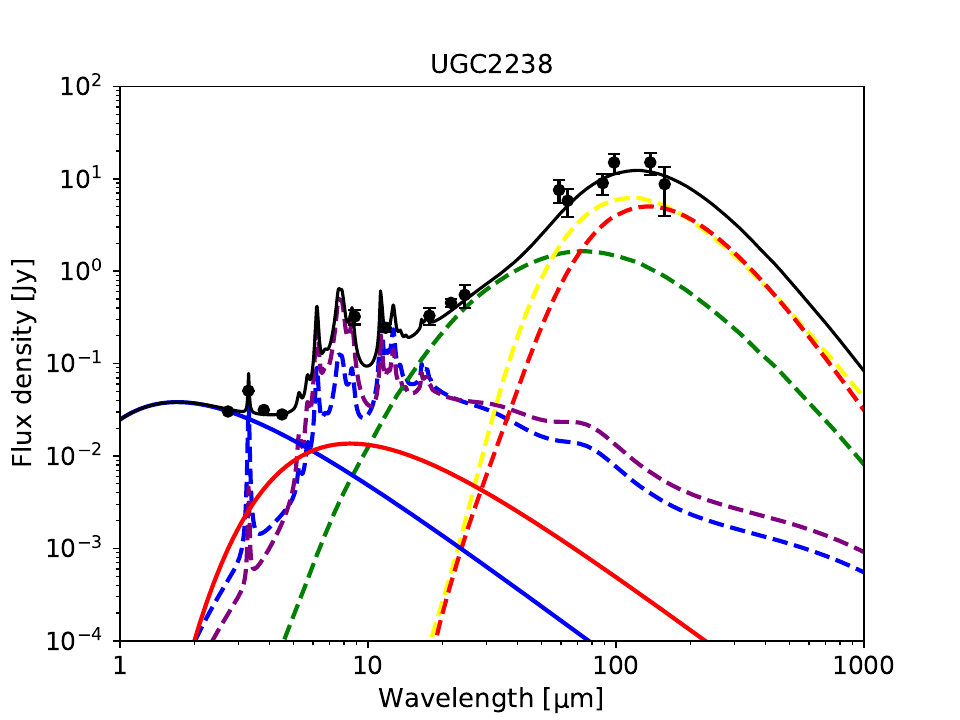} &
\includegraphics[bb=5 0 430 325, clip, width=0.25\textwidth]{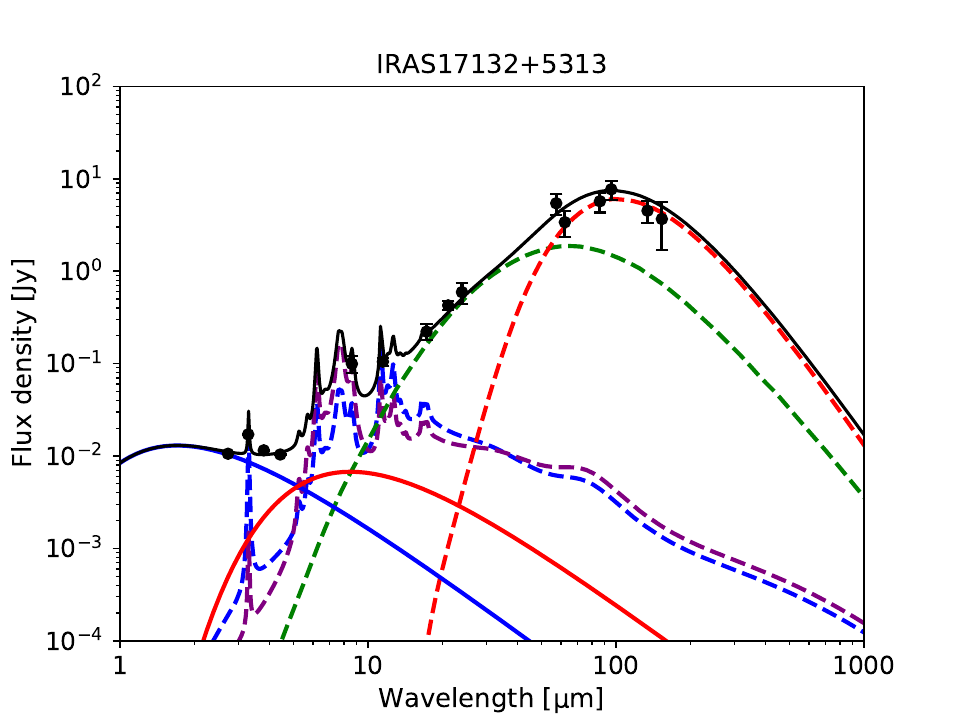} &
\includegraphics[bb=5 0 430 325, clip, width=0.25\textwidth]{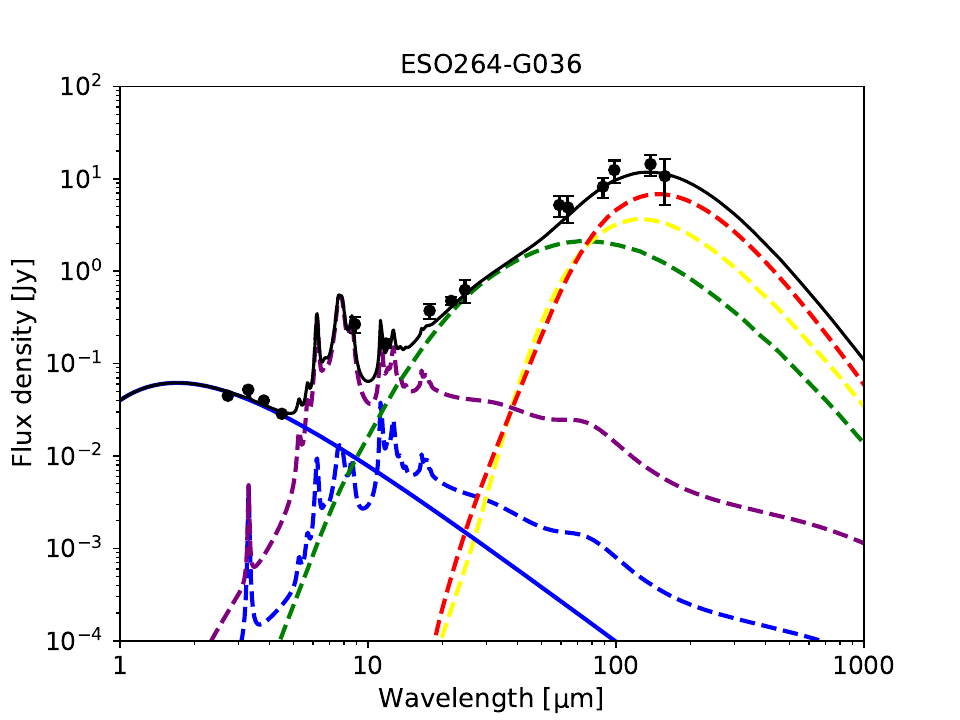} \\
\includegraphics[bb=5 0 430 325, clip, width=0.25\textwidth]{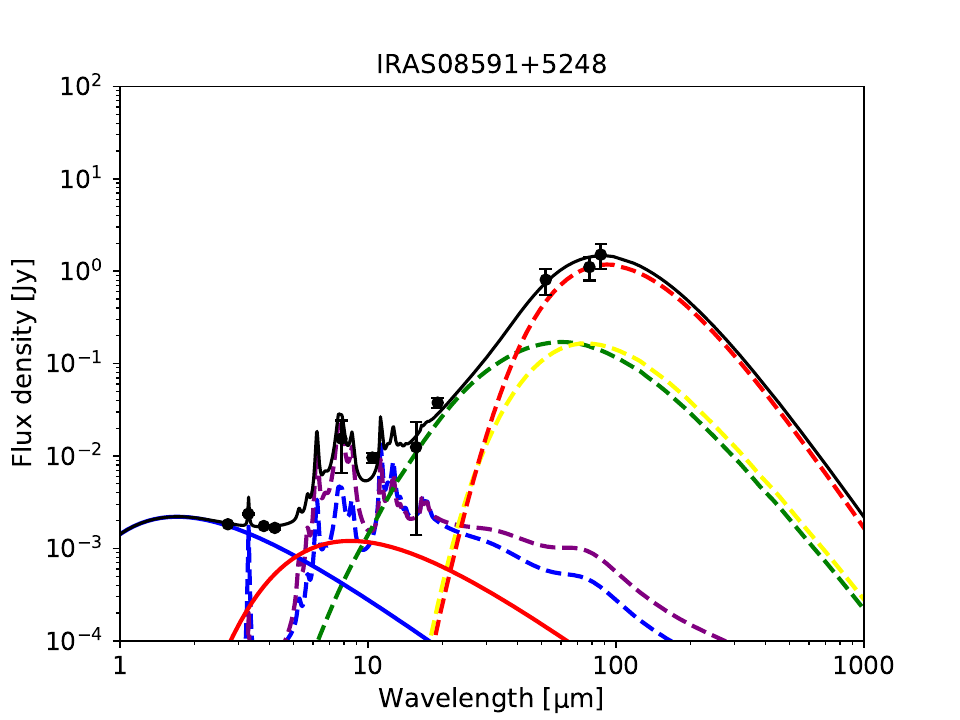} &
\includegraphics[bb=5 0 430 325, clip, width=0.25\textwidth]{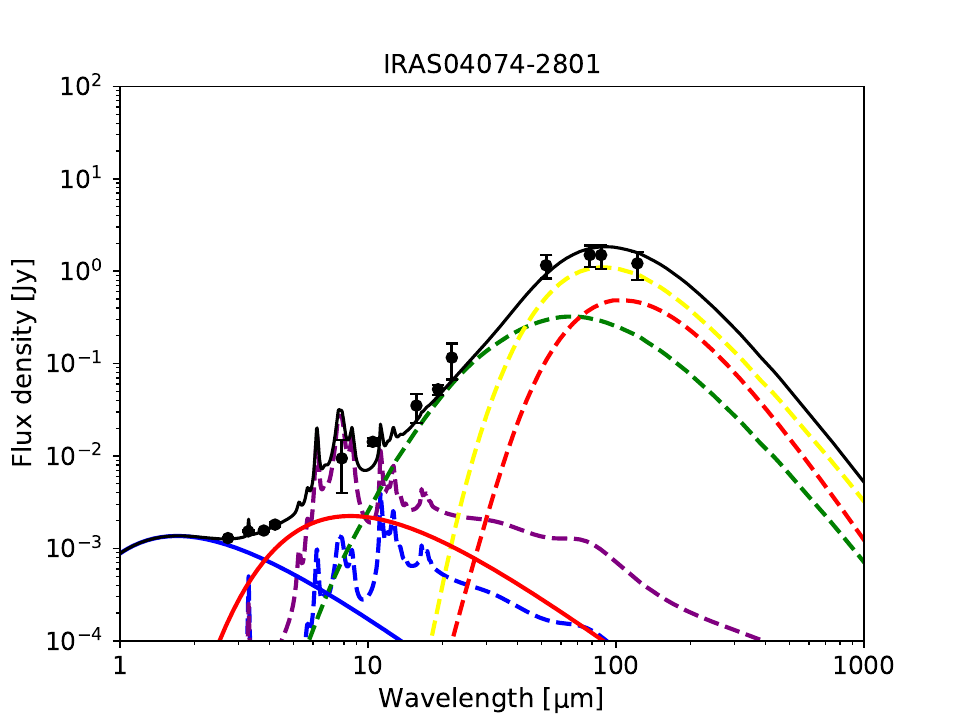} &
\includegraphics[bb=5 0 430 325, clip, width=0.25\textwidth]{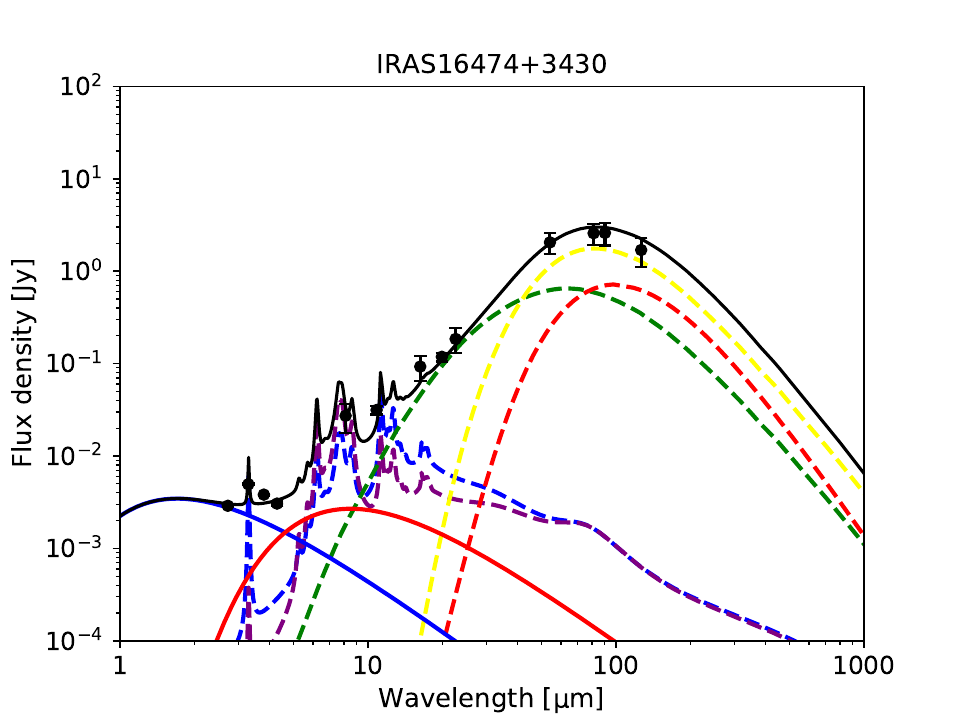} \\
\end{tabular}
\caption{Examples of the SED fitting to the same galaxies as shown in figure \ref{example_nir_fitting}. The SED model is composed of stellar continuum (blue solid line), hot dust continuum (red solid line), ionized PAHs (purple dashed line), neutral PAHs (blue dashed line), small amorphous carbon (green dashed line), large amorphous carbon (yellow dashed line), and amorphous silicate dust (red dashed line). {Alt text: Twelve panels of examples for spectral energy distribution fitting.}}
\label{example_sed_fitting}
\end{figure*}

We perform model fitting to the spectral energy distributions (SEDs) which are composed of near- to far-IR photometric data obtained with AKARI (AKARI All-Sky Survey: IRC Point Source Catalogue Version 1.0; \cite{ishihara2010}, AKARI All-Sky Survey: FIS Bright Source Catalogue Version 2.0, AKARI Far-infrared All-Sky Survey Maps; \cite{doi2015}), WISE (WISE All-sky Source Catalog; \cite{wright2010}), and IRAS (IRAS Faint Source Catalog v2.0; \cite{moshir1990}). Based on the AKARI spectroscopic observation coordinates, the one-to-one source cross-matching is performed for each point source catalog independently with the following search radius: 1.5 arcmin for AKARI/IRC, FIS, WISE and 3.3 arcmin for IRAS, where we do not consider any possible nearby source contamination. In addition to the photometric data, we add to the SEDs four near-IR data points by binning the AKARI near-IR spectra for the wavelength ranges of 2.55--2.90, 3.20--3.38, 3.60--4.00, and 4.10--4.90~\textmu m, following the procedure in \citet{kondo2024}. \\
\indent We use an SED model based on DustEM (\cite{compiegne2011}) which consists of the following three dust components: PAHs, amorphous carbon (amC), and amorphous silicate (aSil). PAHs are composed of neutral and ionized PAHs. We fit the SEDs of the sample galaxies on the condition that the abundance ratio of the neutral to ionized PAHs is free. The amC is divided into large amC (LamC) and small amC (SamC) components based on the dust size. Furthermore, we add two blackbody models with temperatures of 600 and 3000~K, which represent a hot dust and a stellar continuum, respectively. Finally, we perform the SED fitting to the 411 sample galaxies which have more photometric data points than the number of the free parameters of the SED model. Figure \ref{example_sed_fitting} shows examples of the SED fitting results. We estimate the hot dust continuum luminosities, $L_\mathrm{hot}$, the stellar continuum luminosities, $L_\mathrm{star}$, and the total IR luminosities, $L_\mathrm{IR}$ ($=L_\mathrm{star}+L_\mathrm{hot}+L_\mathrm{PAH}+L_\mathrm{amC}+L_\mathrm{aSil}$), of the sample galaxies, which are integrated over the wavelength range of 1--1000~\textmu m.

\begin{figure*}[htbp]
\centering
\begin{tabular}{rcl}
\begin{minipage}[c]{0.32\linewidth}
\centering
\includegraphics[bb=55 0 380 310, clip, width=55mm]{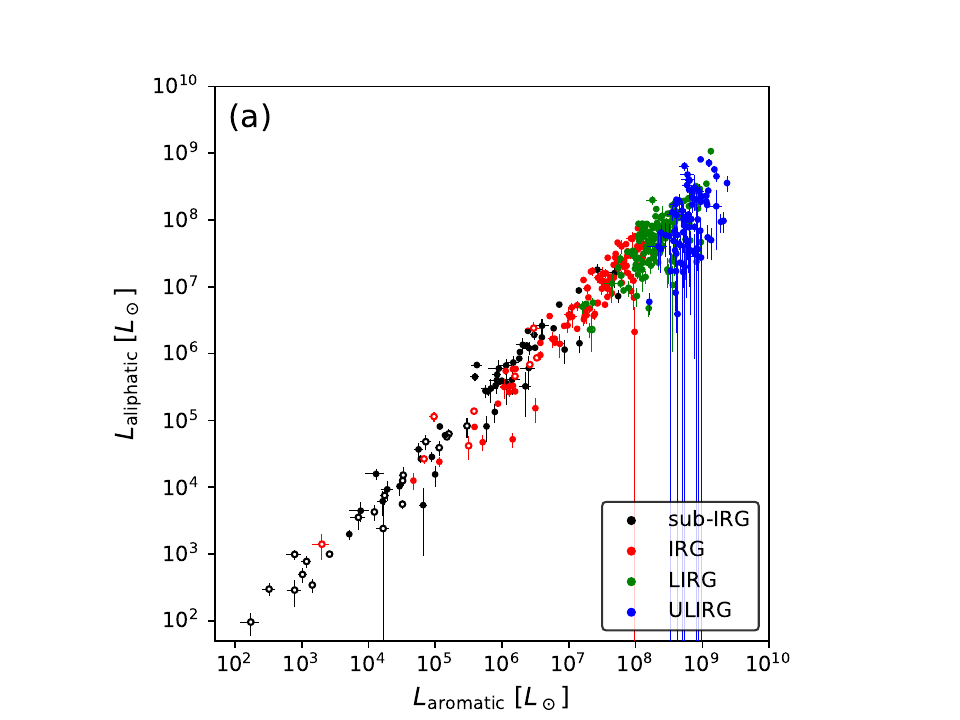}
\end{minipage}
\begin{minipage}[c]{0.32\linewidth}
\centering
\includegraphics[bb=55 0 380 310, clip, width=55mm]{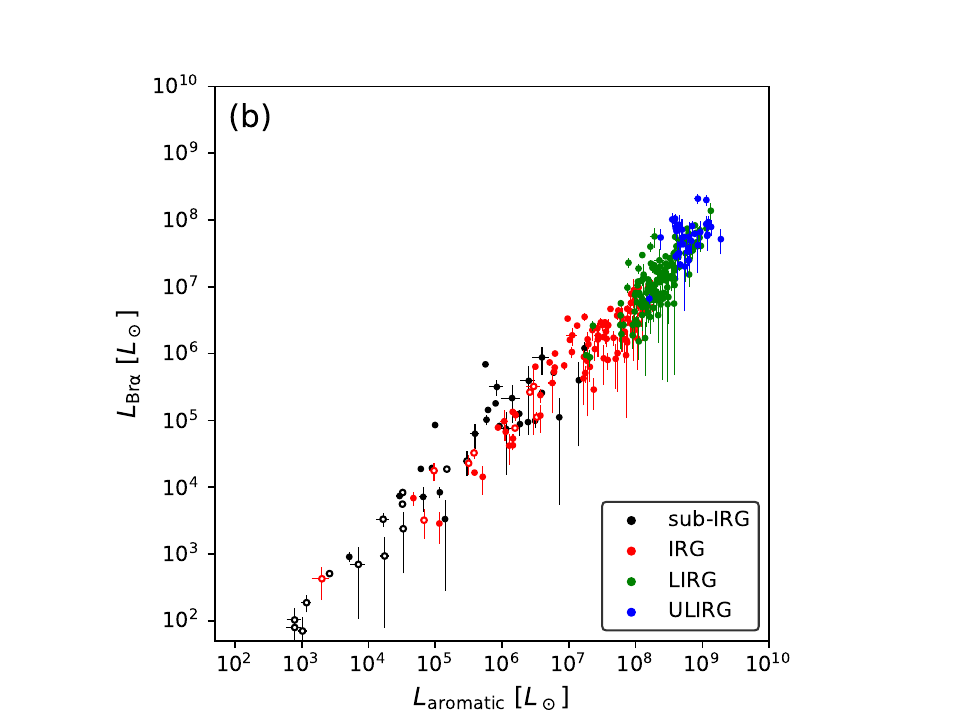}
\end{minipage}
\begin{minipage}[c]{0.32\linewidth}
\centering
\includegraphics[bb=55 0 380 310, clip, width=55mm]{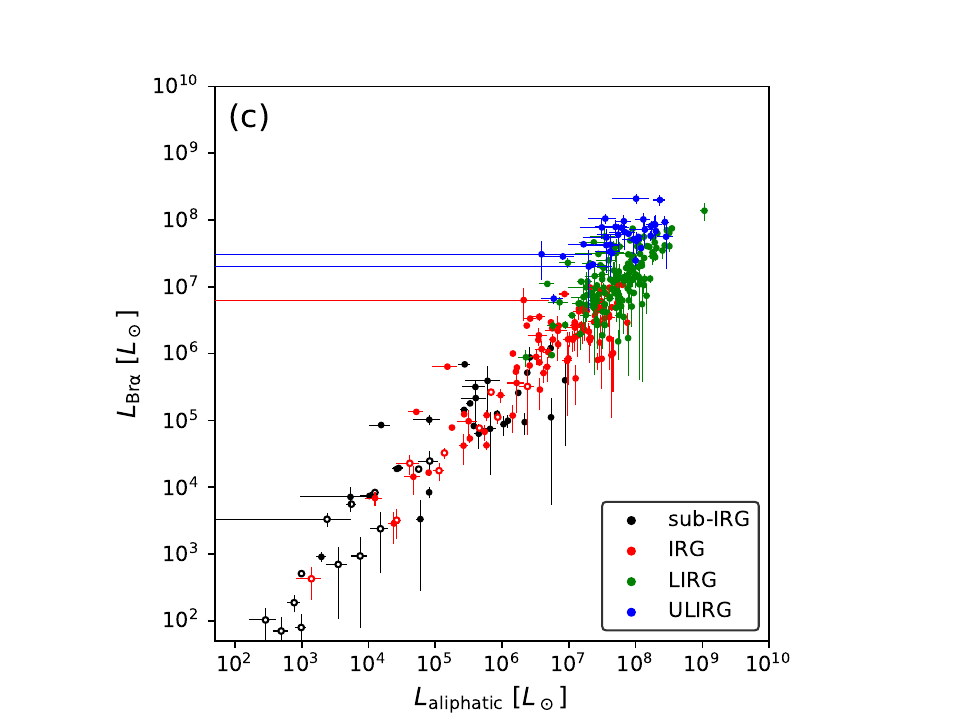}
\end{minipage}
\end{tabular}
\caption{Scatter plots between (a) $L_\mathrm{aliphatic}$ and $L_\mathrm{aromatic}$, (b) $L_\mathrm{Br\alpha}$ and $L_\mathrm{aromatic}$, and (c) $L_\mathrm{Br\alpha}$ and $L_\mathrm{aliphatic}$ for all the sample galaxies observed with $S/N>$ 3 for the detection of the aromatic feature at 3.3~\textmu m and with $S/N>$ 1 for the detection of the aliphatic features at 3.4--3.6~\textmu m and the H\,\emissiontype{I} Br$\mathrm{\alpha}$ recombination line for a display purpose: sub-IRGs (black circles), IRGs (red circles), LIRGs (green circles), and ULIRGs (blue circles). The open marks correspond to the results obtained for off-center regions of a galaxy. {Alt text: Three scatter plots.}}
\label{feature_line_3sigma}
\end{figure*}

\section{Result}
As a result of the near-IR spectral fitting, 428 of the 456 spectra of all the sample galaxies show detection of the aromatic hydrocarbon feature at 3.3~\textmu m with $S/N>$ 3, the breakdown of which is 239 for (U)LIRGs, 117 for IRGs, and 72 for sub-IRGs. Figure \ref{feature_line_3sigma} shows correlation plots between $L_\mathrm{aromatic}$, $L_\mathrm{aliphatic}$, and $L_\mathrm{Br\alpha}$ estimated from the 456 spectra, from which it is confirmed that those parameters are significantly correlated with each other. Notably, the present study covers as wide a range of $L_\mathrm{aromatic}$ as about 8 orders of magnitude ($10^{2}$--$10^{10}\ \mathrm{L_\odot}$). On the other hand, \citet{kondo2024} covered a range of $10^{6}$--$10^{10}\ \mathrm{L_\odot}$, only about a higher half of the luminosity range covered by the present study.

\subsection{Variations in $L_\mathrm{aliphatic}/L_\mathrm{aromatic}$ from galaxy to galaxy}
\begin{figure*}[htbp]
\centering
\includegraphics[bb=5 5 415 305, clip, width=100mm]{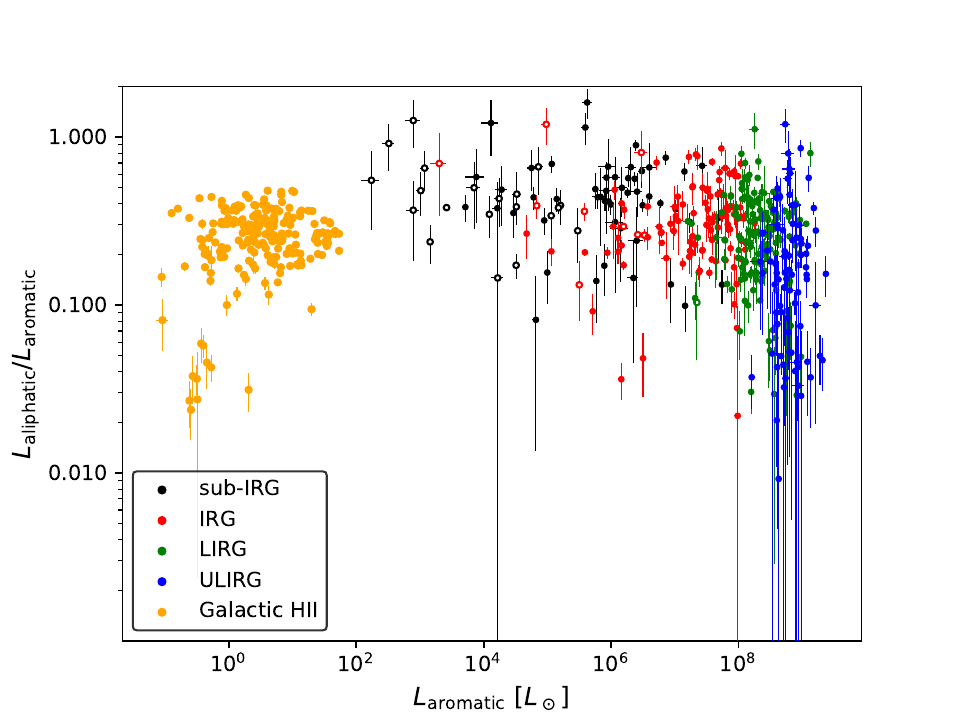}
\caption{$L_\mathrm{aliphatic}/L_\mathrm{aromatic}$ plotted against $L_\mathrm{aromatic}$ for all the sample galaxies observed with $S/N>$ 3 for the detection of the aromatic feature at 3.3~\textmu m: sub-IRGs (black circles), IRGs (red circles), LIRGs (green circles), ULIRGs (blue circles), and Galactic H\,\emissiontype{II} regions (orange circles). The open marks correspond to the results obtained for off-center regions of a galaxy. {Alt text: One scatter plot.}}
\label{aliaro_laro_3sigma}
\end{figure*}

\begin{figure*}[htbp]
\centering
\begin{tabular}{rl}
\begin{minipage}[c]{0.45\linewidth}
\centering
\includegraphics[bb=48 0 380 305, clip, width=0.9\linewidth]{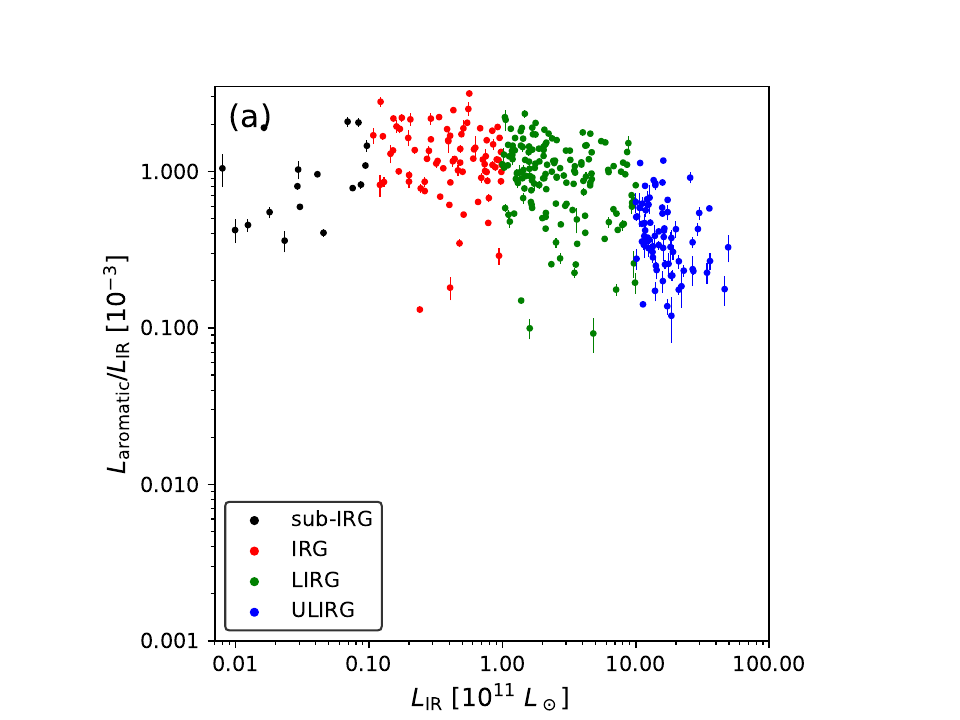}
\end{minipage}
\begin{minipage}[c]{0.45\linewidth}
\centering
\includegraphics[bb=48 0 380 305, clip, width=0.9\linewidth]{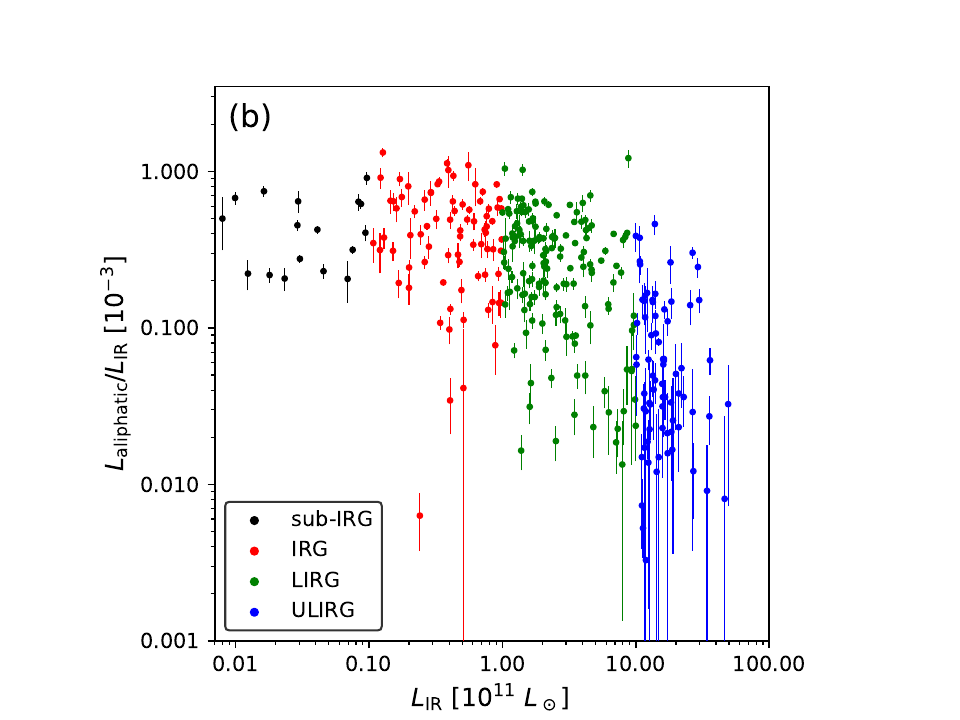}
\end{minipage} \\
\multicolumn{2}{c}{
    \begin{minipage}[c]{0.9\linewidth}
    \centering
    \includegraphics[bb=50 0 380 305, clip, width=0.45\linewidth]{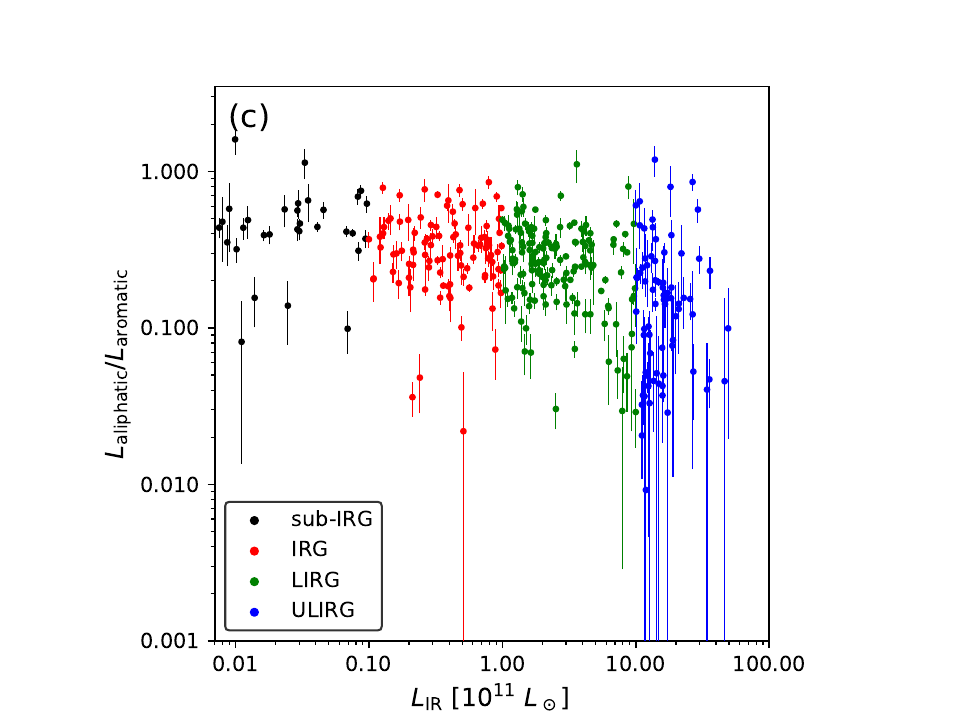}
    \end{minipage}
    }
\end{tabular}
\caption{(a) $L_\mathrm{aromatic}/L_\mathrm{IR}$, (b) $L_\mathrm{aliphatic}/L_\mathrm{IR}$, and (c) $L_\mathrm{aliphatic}/L_\mathrm{aromatic}$ all as a function of $L_\mathrm{IR}$ for the sample galaxies observed with $S/N>$ 3 for the detection of the aromatic feature at 3.3~\textmu m: sub-IRG (black circles), IRGs (red circles), LIRGs (green circles), and ULIRGs (blue circles). {Alt text: Three scatter plots.}}
\label{aliaro_lir_3sigma}
\end{figure*}

Figure \ref{aliaro_laro_3sigma} shows $L_\mathrm{aliphatic}/L_\mathrm{aromatic}$ plotted against $L_\mathrm{aromatic}$ for all the sample galaxies. As a reference, we overplot the data points of the Galactic H\,\emissiontype{II} regions. As seen in the figure, the values of $L_\mathrm{aliphatic}/L_\mathrm{aromatic}$ of the sample galaxies are widely distributed from 0.01 to 1, which exhibit considerably large variations, compared to those of the Galactic H\,\emissiontype{II} regions. In addition, $L_\mathrm{aliphatic}/L_\mathrm{aromatic}$ is systematically different across the luminosity classes. The median values of $L_\mathrm{aliphatic}/L_\mathrm{aromatic}$ for the sub-IRGs, the IRGs, the LIRGs, the ULIRGs, and the Galactic H\,\emissiontype{II} regions are 0.43, 0.30, 0.26, 0.15, and 0.26, respectively, and thus the LIRGs show $L_\mathrm{aliphatic}/L_\mathrm{aromatic}$ values similar to those of the Galactic H\,\emissiontype{II} regions. On the other hand, the $L_\mathrm{aliphatic}/L_\mathrm{aromatic}$ values of the (sub-)IRGs are systematically higher, while the $L_\mathrm{aliphatic}/L_\mathrm{aromatic}$ values of the ULIRGs are systematically lower than those of the Galactic H\,\emissiontype{II} regions. These results imply that the properties of hydrocarbon dust in the (sub-)IRGs and the ULIRGs somewhat differ from those in the LIRGs and the Galactic H\,\emissiontype{II} regions. \\
\indent We plot $L_\mathrm{aromatic}/L_\mathrm{IR}$, $L_\mathrm{aliphatic}/L_\mathrm{IR}$, and $L_\mathrm{aliphatic}/L_\mathrm{aromatic}$ as a function of $L_\mathrm{IR}$ in figures \ref{aliaro_lir_3sigma}a, \ref{aliaro_lir_3sigma}b, and \ref{aliaro_lir_3sigma}c, respectively. As already pointed out by \citet{kondo2024}, $L_\mathrm{aromatic}/L_\mathrm{IR}$, $L_\mathrm{aliphatic}/L_\mathrm{IR}$, and $L_\mathrm{aliphatic}/L_\mathrm{aromatic}$ decrease with $L_\mathrm{IR}$ at higher $L_\mathrm{IR}\ (>10^{11}\ \mathrm{L_\odot})$. Owing to improved statistics in IRG and sub-IRG sample in the present study, figure \ref{aliaro_lir_3sigma}c shows that $L_\mathrm{aliphatic}/L_\mathrm{aromatic}$ decreases systematically with $L_\mathrm{IR}$ over the range of $10^9$--$10^{13}\ \mathrm{L_\odot}$, while this trend was recognized only at higher $L_\mathrm{IR}\ (>10^{11}\ \mathrm{L_\odot})$ in \citet{kondo2024}.

\begin{figure*}[htbp]
\centering
\begin{tabular}{rl}
\begin{minipage}[c]{0.45\linewidth}
\centering
\includegraphics[bb=50 0 380 305, clip, width=0.9\linewidth]{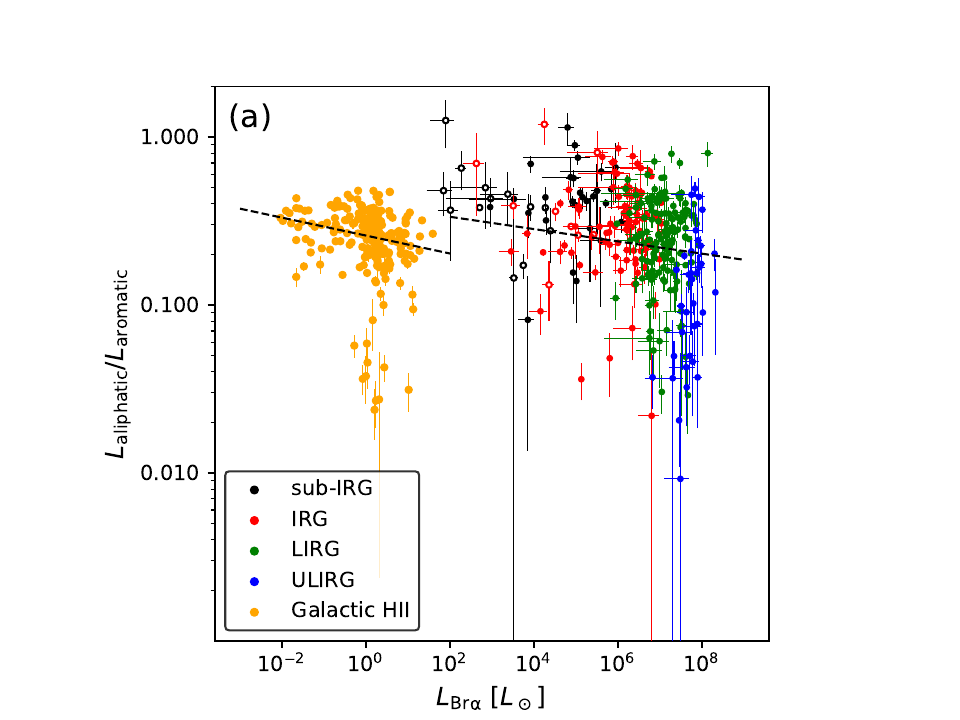}
\end{minipage}
\begin{minipage}[c]{0.45\linewidth}
\centering
\includegraphics[bb=50 0 380 305, clip, width=0.9\linewidth]{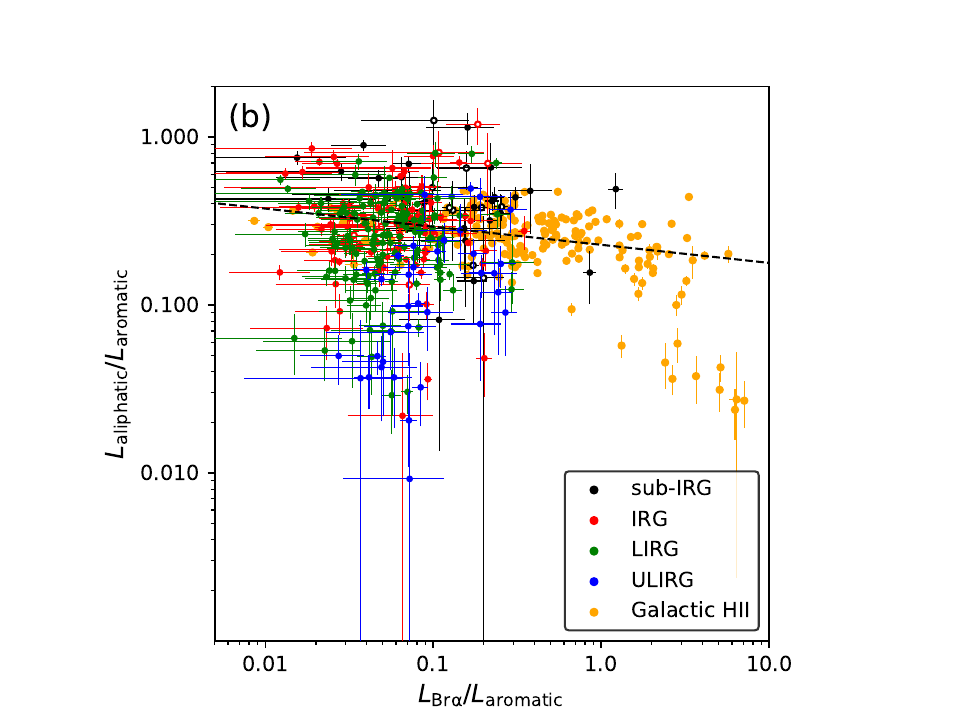}
\end{minipage}
\end{tabular}
\caption{$L_\mathrm{aliphatic}/L_\mathrm{aromatic}$ as functions of (a) $L_\mathrm{Br\alpha}$ and (b) $L_\mathrm{Br\alpha}/L_\mathrm{aromatic}$ for all the sample observed with $S/N>$ 3 for the detection of the aromatic feature at 3.3~\textmu m and with $S/N>$ 1 for the detection of the aliphatic features at 3.4--3.6~\textmu m and the H\,\emissiontype{I} Br$\mathrm{\alpha}$ recombination line for a display purpose: sub-IRG (black circles), IRGs (red circles), LIRGs (green circles), ULIRGs (blue circles), and Galactic H\,\emissiontype{II} regions (orange circles). The open marks correspond to the results obtained for off-center regions of a galaxy. The dashed lines in panel(a) are the ones fitted to the sample galaxies and the Galactic H\,\emissiontype{II} regions. In contrast, the dashed line in panel(b) is the one fitted to the sample galaxies alone. {Alt text: Two scatter plots.}}
\label{aliaro_bra_3sigma}
\end{figure*}

\subsection{Relationship between $L_\mathrm{aliphatic}/L_\mathrm{aromatic}$ and the interstellar radiation field}
\label{aliaro_isrf}
\citet{mori2014} and \citet{kondo2024} showed that $L_\mathrm{aliphatic}/L_\mathrm{aromatic}$ decreases with the strength of the UV radiation field. \citet{mori2014} used $I_\mathrm{cont,\ 3.7\ \mu m}/I_\mathrm{aromatic}$, which is considered to reflect the ionization degree of PAHs (\cite{haraguchi2012}) and thus as an indicator of the UV radiation field strength. \citet{kondo2024} used $G_0$, which is defined by the far-UV (6.0--13.6~eV) radiation field strength normalized to the solar neighborhood value (\cite{habing1968}). In the case of galaxies, the interpretation of $I_\mathrm{cont,\ 3.7\ \mu m}/I_\mathrm{aromatic}$ is not straightforward because $I_\mathrm{cont,\ 3.7\ \mu m}$ generally traces not only ionized PAHs but also stellar components. Also, we cannot calculate $G_0$ for some galaxies, for which we lack the far-IR data needed to derive $G_0$ from the SED fitting. Therefore, in the present study, we use $L_\mathrm{Br\alpha}$ as a tracer of the UV radiation fields, which traces H\,\emissiontype{II} regions exposed to extreme-UV radiation ($>$ 13.6~eV). Figure \ref{aliaro_bra_3sigma}a shows $L_\mathrm{aliphatic}/L_\mathrm{aromatic}$ as a function of $L_\mathrm{Br\alpha}$, where we confirm that both galaxies and Galactic H\,\emissiontype{II} regions show gradually decreasing trends with $L_\mathrm{Br\alpha}$, which is consistent with the results of \citet{mori2014} and \citet{kondo2024}. \\
\indent We also examine the dependence of $L_\mathrm{aliphatic}/L_\mathrm{aromatic}$ on the radiation hardness. Here we define the hardness by $L_\mathrm{Br\alpha}/L_\mathrm{aromatic}$ since $L_\mathrm{aromatic}$ emission is mostly associated with the far-UV while $L_\mathrm{Br\alpha}$ with the extreme-UV. As seen in figure \ref{aliaro_bra_3sigma}b, $L_\mathrm{aliphatic}/L_\mathrm{aromatic}$ globally decreases with $L_\mathrm{Br\alpha}/L_\mathrm{aromatic}$ and thus the radiation hardness, where there seems to be no systematic difference between the galaxies and the Galactic H\,\emissiontype{II} regions. The scatter of the galaxies, however, is considerably larger, and in particular, some of the ULIRGs are largely deviated from the global trend, showing extremely low $L_\mathrm{aliphatic}/L_\mathrm{aromatic}$ ($<$ 0.1), possible causes of which will be discussed in section \ref{aliaro_high_luminosity}. Hence we cannot explain the variations of $L_\mathrm{aliphatic}/L_\mathrm{aromatic}$ by the effect of the radiation hardness alone.

\section{Discussion}
\subsection{Robustness of the near-IR fitting results}
\label{robustness_fitting}
\begin{figure}[htbp]
\centering
\includegraphics[bb=65 5 370 320, width=0.825\linewidth]{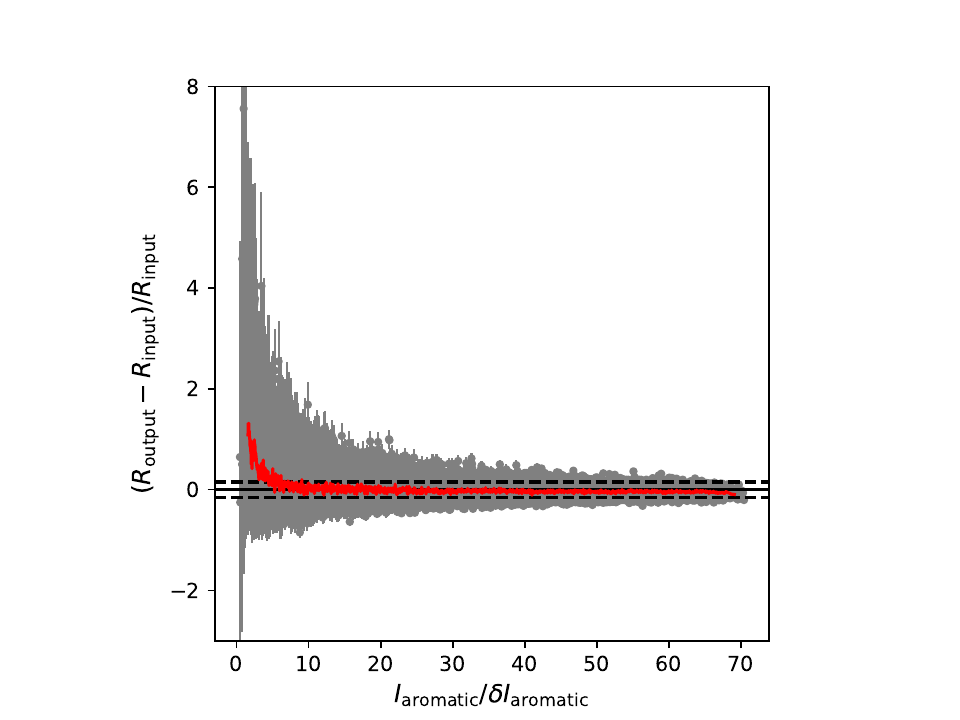}
\caption{Dependence of the fitting result of $R_\mathrm{output}=L_\mathrm{aliphatic}/L_\mathrm{aromatic}$ on the $S/N$ based on the simulated spectral data. The red curve shows median values calculated per 51 points along the horizontal axis. The dashed lines represent $(R_\mathrm{output}-R_\mathrm{input})/R_\mathrm{input}=\pm$ 0.15. {Alt text: One scatter plot.}}
\label{simulation}
\end{figure}

\begin{figure*}[htbp]
\centering
\begin{tabular}{rl}
\begin{minipage}[c]{0.45\linewidth}
\centering
\includegraphics[bb=50 0 380 305, clip, width=0.9\linewidth]{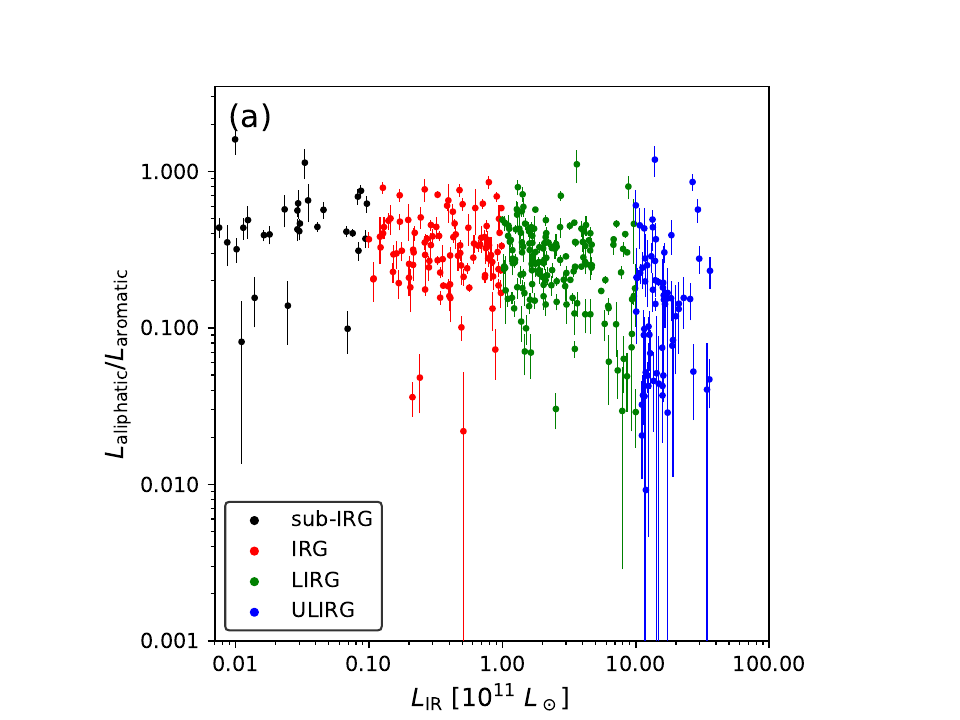}
\end{minipage}
\begin{minipage}[c]{0.45\linewidth}
\centering
\includegraphics[bb=50 0 380 305, clip, width=0.9\linewidth]{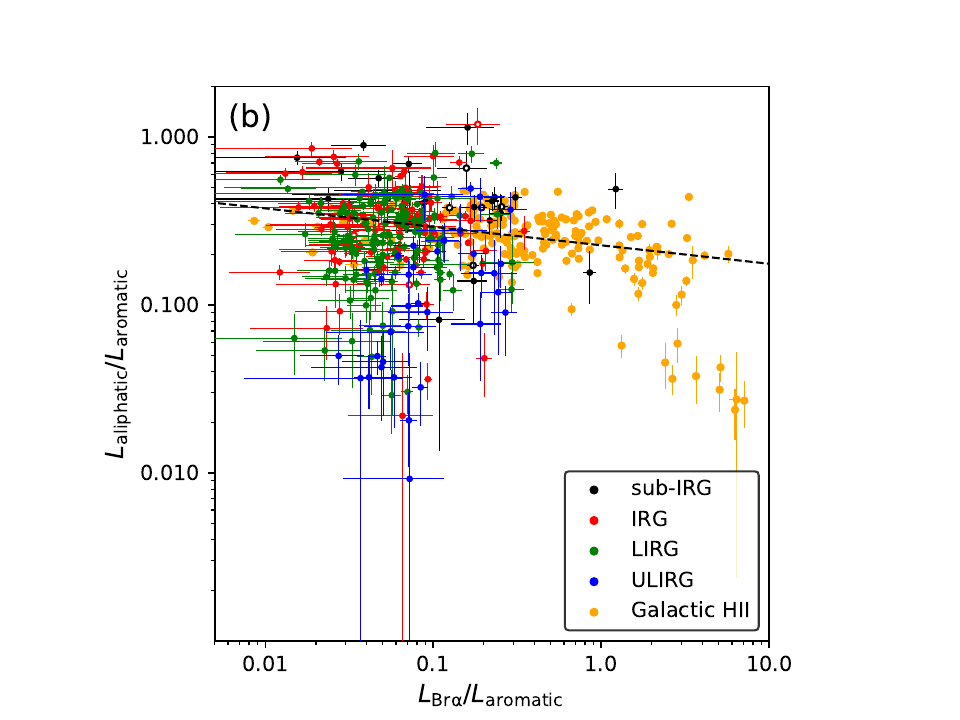}
\end{minipage}
\end{tabular}
\caption{(a) Same as figure \ref{aliaro_lir_3sigma}c and (b) same as figure \ref{aliaro_bra_3sigma}b, but only data points with $I_\mathrm{aromatic}/\delta I_\mathrm{aromatic}\geq$ 5.0 are shown. {Alt text: Two scatter plots.}}
\label{aliaro_5sigma}
\end{figure*}

To check the robustness of the model fitting, we investigate the dependence of the fitting results on the $S/N$ of the fitted data. Based on the typical model spectrum of our sample galaxies with $L_\mathrm{aliphatic}/L_\mathrm{aromatic}=$ 0.3, we generate Gaussian errors with various levels of $S/N$ randomly and add them to the model spectrum. Then, we perform the model fitting to 30000 simulated spectra thus created for a range of $S/N=$ 1 to 30 with a step of 0.001. Figure \ref{simulation} shows $(R_\mathrm{output}-R_\mathrm{input})/R_\mathrm{input}$, where $R_\mathrm{input}$ is the input $L_\mathrm{aliphatic}/L_\mathrm{aromatic}$ value of the model spectrum (i.e., $R_\mathrm{input}=$ 0.3) and $R_\mathrm{output}$ is the best-fit value of $L_\mathrm{aliphatic}/L_\mathrm{aromatic}$, as a function of $I_\mathrm{aromatic}/\delta I_\mathrm{aromatic}$ estimated from the spectral fitting. Figure \ref{simulation} indicates that the systematics become significant and the robustness can be lost when $I_\mathrm{aromatic}/\delta I_\mathrm{aromatic}$ is lower than 5.0. Hence, in the following discussion, we select only the spectral data with the fitting results which satisfy the threshold of $I_\mathrm{aromatic}/\delta I_\mathrm{aromatic}\geq$ 5.0. Figures \ref{aliaro_5sigma}a and \ref{aliaro_5sigma}b are the same as figures \ref{aliaro_lir_3sigma}c and \ref{aliaro_bra_3sigma}b, but only data points with $I_\mathrm{aromatic}/\delta I_\mathrm{aromatic}\geq$ 5.0 are shown, from which we confirm that the trends in figures \ref{aliaro_lir_3sigma}c and \ref{aliaro_bra_3sigma}b do not change significantly with $S/N\geq$ 5.

\subsection{Dependence of $L_\mathrm{aliphatic}/L_\mathrm{aromatic}$ on near-IR continuum properties}
\label{aliaro_galaxy_property}

\begin{figure*}[htbp]
\centering
\begin{tabular}{rl}
\begin{minipage}[c]{0.45\linewidth}
\centering
\includegraphics[bb=50 0 380 305, clip, width=0.9\linewidth]{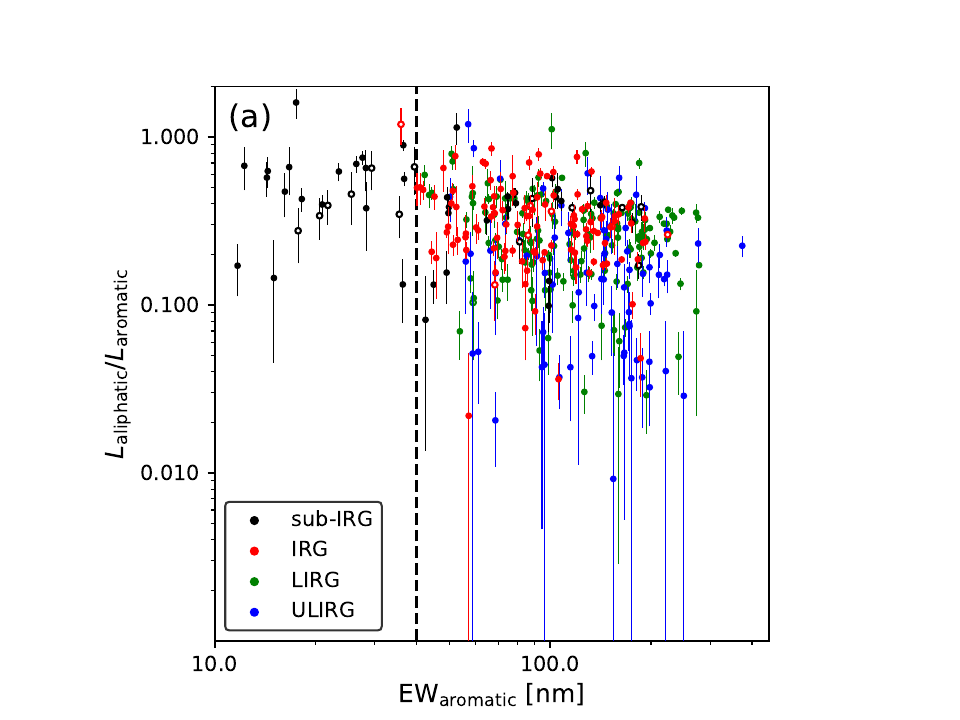}
\end{minipage}
\begin{minipage}[c]{0.45\linewidth}
\centering
\includegraphics[bb=50 0 380 305, clip, width=0.9\linewidth]{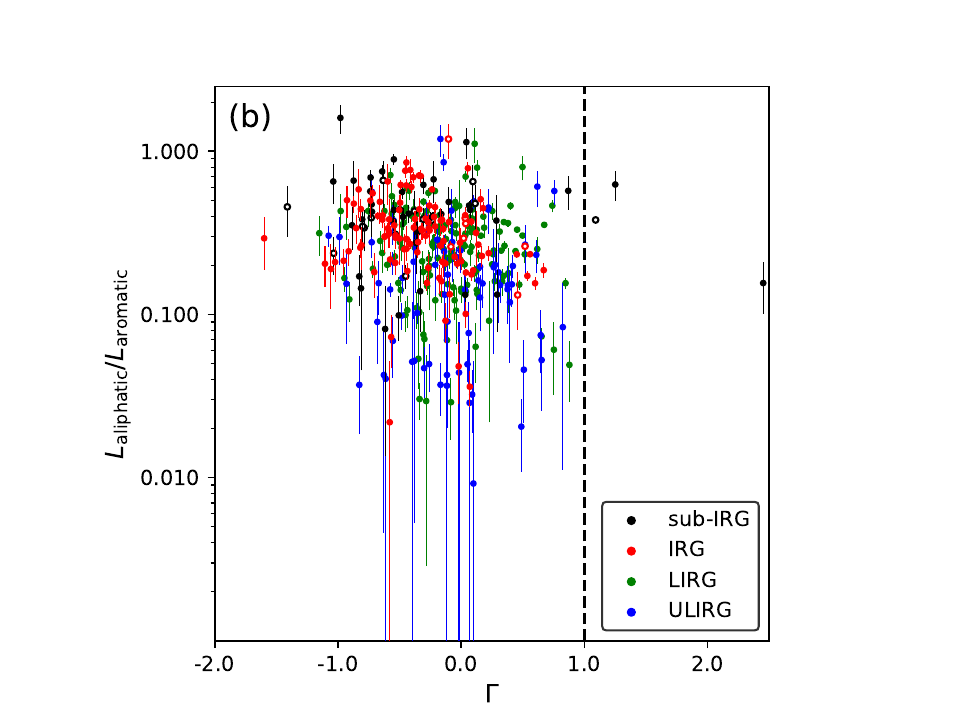}
\end{minipage} \\
\multicolumn{2}{c}{
    \begin{minipage}[c]{0.9\linewidth}
    \centering
    \includegraphics[bb=50 0 380 305, clip, width=0.45\linewidth]{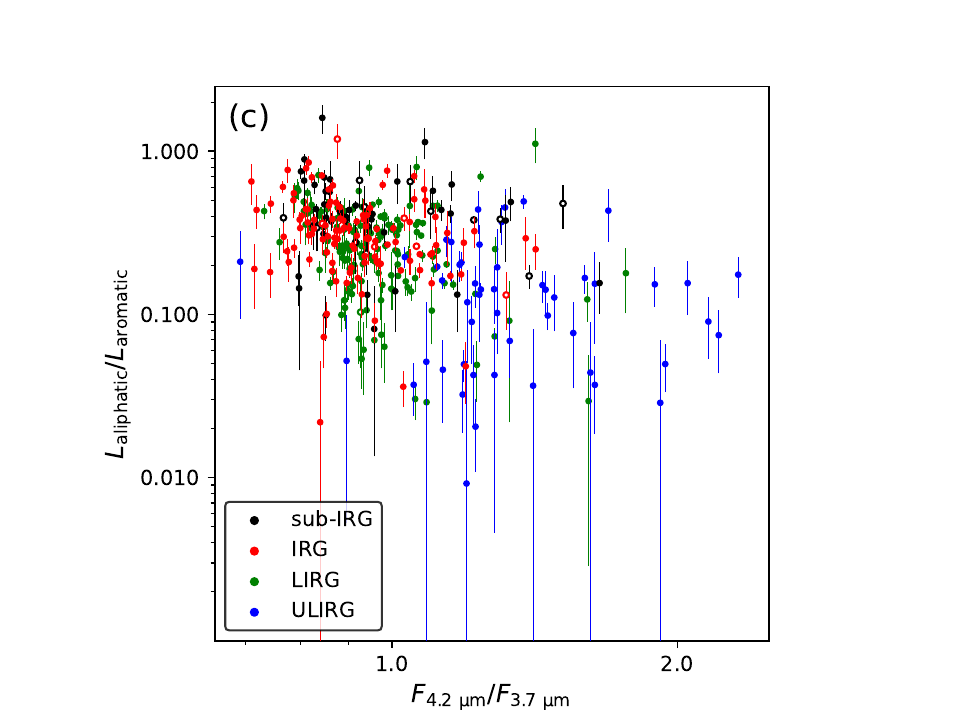}
    \end{minipage}
    }
\end{tabular}
\caption{$L_\mathrm{aliphatic}/L_\mathrm{aromatic}$ as functions of (a) $\mathrm{EW_{aromatic}}$, (b) $\Gamma$, and (c) $F_\mathrm{4.2\ \mu m}/F_\mathrm{3.7\ \mu m}$ for the sample galaxies, color-coded with the luminosity class: sub-IRG (black circles), IRGs (red circles), LIRGs (green circles), and ULIRGs (blue circles). The open marks correspond to the results obtained for off-center regions of a galaxy. The dashed lines in panels(a) and (b) correspond to $\mathrm{EW_{aromatic}}=$ 40~nm and $\Gamma=$ 1, respectively, which are used as thresholds to identify the pure star-forming (U)LIRG and IRG samples (i.e., $\mathrm{EW_{aromatic}}>$ 40~nm and $\Gamma<$ 1). {Alt text: Three scatter plots.}}
\label{aliaro_luminosityclass}
\end{figure*}

First, we investigate whether there is any dependence of $L_\mathrm{aliphatic}/L_\mathrm{aromatic}$ on $\mathrm{EW_{aromatic}}$ and $\Gamma$, the parameters used for the galaxy classification in section \ref{data_cls_selection}. Figures \ref{aliaro_luminosityclass}a and \ref{aliaro_luminosityclass}b show $L_\mathrm{aliphatic}/L_\mathrm{aromatic}$ as functions of $\mathrm{EW_{aromatic}}$ and $\Gamma$, respectively, from which we confirm that, for each luminosity class, $L_\mathrm{aliphatic}/L_\mathrm{aromatic}$ does not have clear dependence on either $\mathrm{EW_{aromatic}}$ or $\Gamma$, only the systematic differences between the different luminosity classes outstanding. Thus the underlying continuum intensity ($\mathrm{EW_{aromatic}}$) and the 3~\textmu m continuum color ($\Gamma$) are not important parameters to explain the variations of $L_\mathrm{aliphatic}/L_\mathrm{aromatic}$. \\
\indent On the contrary, we find that the 4~\textmu m continuum color ($F_\mathrm{4.2\ \mu m}/F_\mathrm{3.7\ \mu m}$) is a key parameter, as shown in figure \ref{aliaro_luminosityclass}c. Not only within each luminosity class but also over all the luminosity range, $L_\mathrm{aliphatic}/L_\mathrm{aromatic}$ consistently decreases with the 4~\textmu m continuum color from blue to red, where the (sub-)IRGs tend to be bluer while the ULIRGs redder at 4~\textmu m. In the following sections, we discuss what makes the 4~\textmu m continuum bluer and $L_\mathrm{aliphatic}/L_\mathrm{aromatic}$ higher for the (sub-)IRGs, while the 4~\textmu m continuum redder and $L_\mathrm{aliphatic}/L_\mathrm{aromatic}$ lower for the ULIRGs.

\subsection{A possible cause for high $L_\mathrm{aliphatic}/L_\mathrm{aromatic}$ of (sub-)IRGs}
\label{aliaro_low_luminosity}

\begin{figure*}[htbp]
\centering
\begin{tabular}{rl}
\begin{minipage}[c]{0.45\linewidth}
\centering
\includegraphics[bb=50 0 380 310, clip, width=0.85\linewidth]{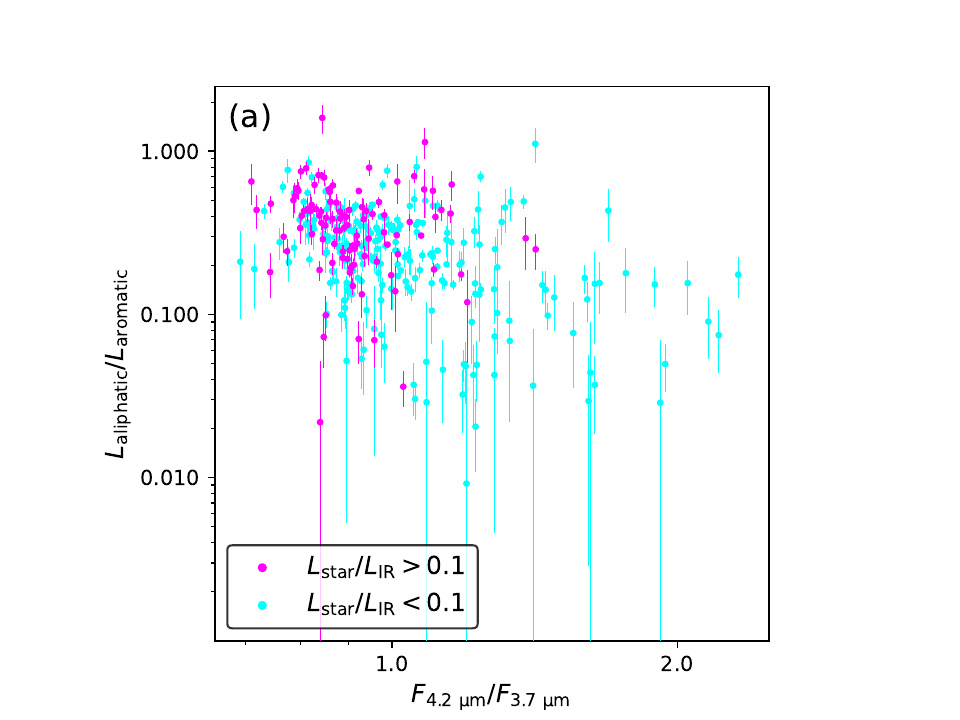}
\end{minipage}
\begin{minipage}[c]{0.45\linewidth}
\centering
\includegraphics[bb=50 0 380 310, clip, width=0.85\linewidth]{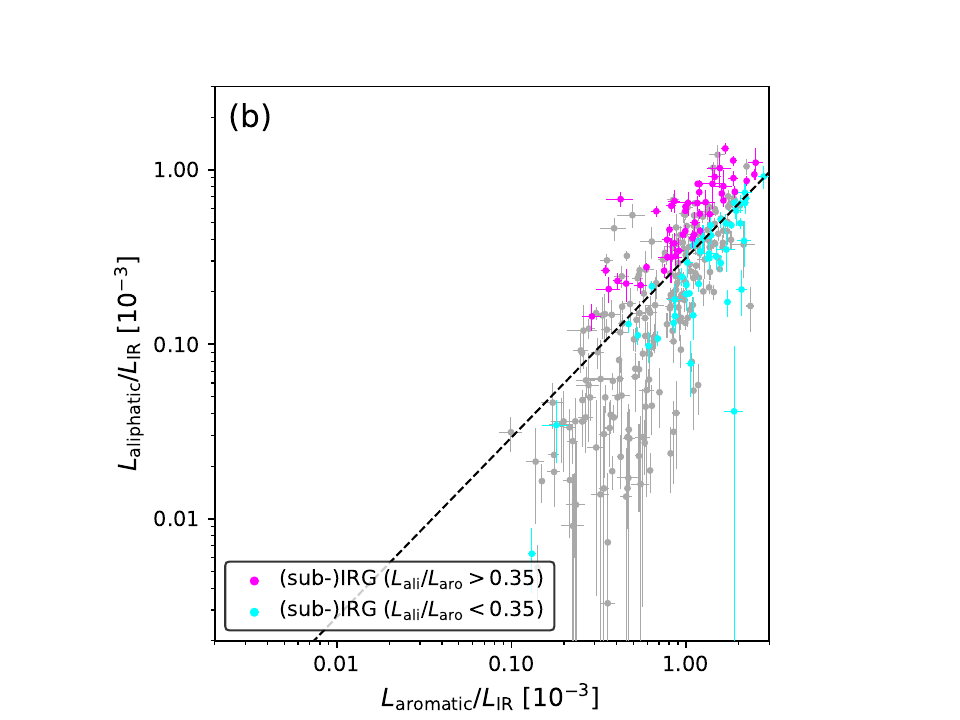}
\end{minipage} \\
\begin{minipage}[c]{0.45\linewidth}
\centering
\includegraphics[bb=50 0 380 310, clip, width=0.85\linewidth]{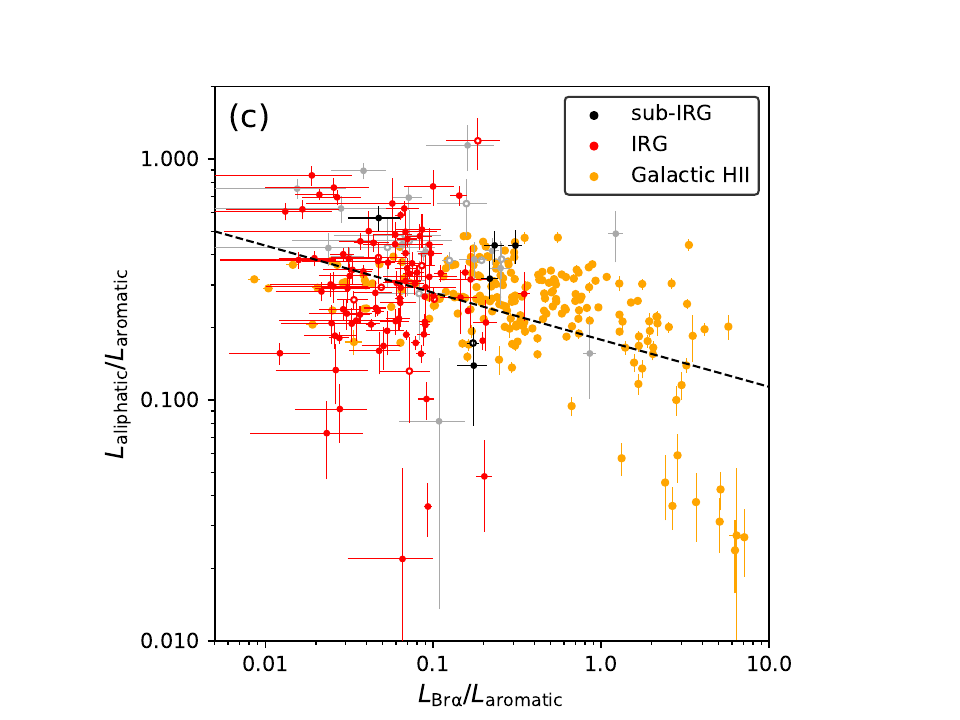}
\end{minipage}
\begin{minipage}[c]{0.45\linewidth}
\centering
\includegraphics[bb=50 0 380 310, clip, width=0.85\linewidth]{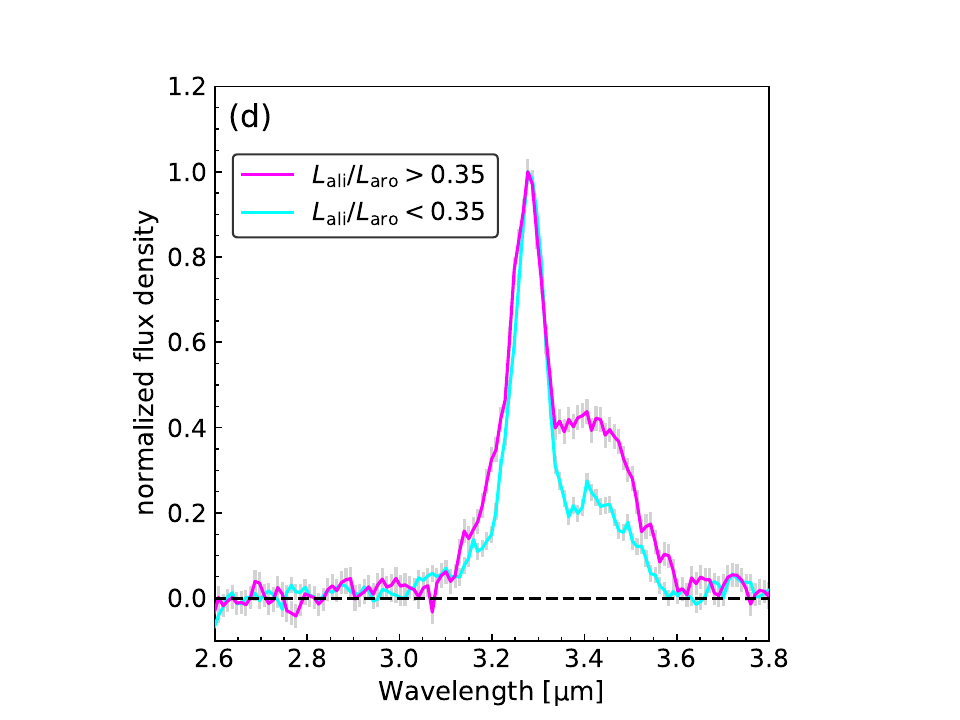}
\end{minipage}
\end{tabular}
\caption{(a) Same as figure \ref{aliaro_luminosityclass}c, but color-coded with $L_\mathrm{star}/L_\mathrm{IR}$: $L_\mathrm{star}/L_\mathrm{IR}>$ 0.1 (magenta circles) and $L_\mathrm{star}/L_\mathrm{IR}<$ 0.1 (cyan circles). (b) Relationship between $L_\mathrm{aliphatic}/L_\mathrm{IR}$ and $L_\mathrm{aromatic}/L_\mathrm{IR}$ for the (sub-)IRG samples: $L_\mathrm{aliphatic}/L_\mathrm{aromatic}>$ 0.35 (magenta circles) and $L_\mathrm{aliphatic}/L_\mathrm{aromatic}<$ 0.35 (cyan circles). The (U)LIRG samples are shown in grey color. The dashed line in panel(b) is the one fitted to the (U)LIRGs data points. (c) Same as figure \ref{aliaro_5sigma}b, but plotted only for the (sub-)IRGs and the Galactic H\,\emissiontype{II} regions. The grey data points represent sub-IRGs which are reported to have AGNs based on NED or SIMBAD\footnote{http://simbad.u-strasbg.fr/simbad/}. The dashed line in panel(c) is the one fitted to the non-AGN (sub-)IRGs data points. (d) The stacked spectra of the non-AGN (sub-)IRG samples with $L_\mathrm{aliphatic}/L_\mathrm{aromatic}>$ 0.35 (magenta) and $L_\mathrm{aliphatic}/L_\mathrm{aromatic}<$ 0.35 (cyan), where the $\mathrm{H_2O}$ ice absorption is corrected and the continuum is subtracted (see equation (\ref{all_model}) in text). {Alt text: Three scatter plots and one graph.}}
\label{aliaro_lstarlir}
\end{figure*}

\begin{figure*}[htbp]
\centering
\begin{tabular}{rl}
\begin{minipage}[c]{0.45\linewidth}
\centering
\includegraphics[bb=50 0 380 305, clip, width=0.9\linewidth]{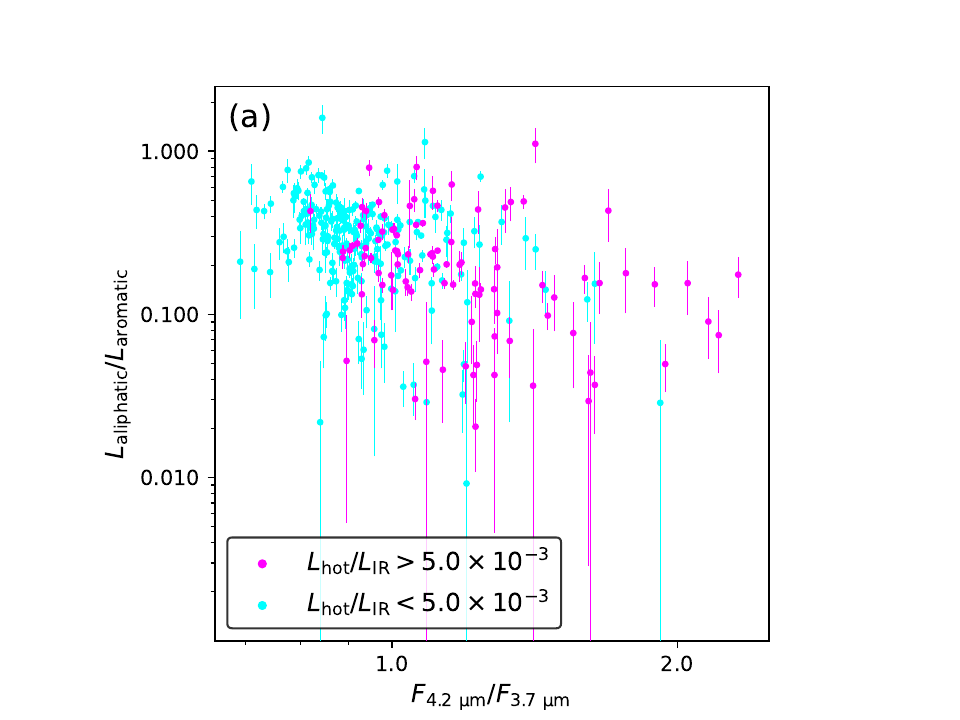}
\end{minipage}
\begin{minipage}[c]{0.45\linewidth}
\centering
\includegraphics[bb=50 0 380 305, clip, width=0.9\linewidth]{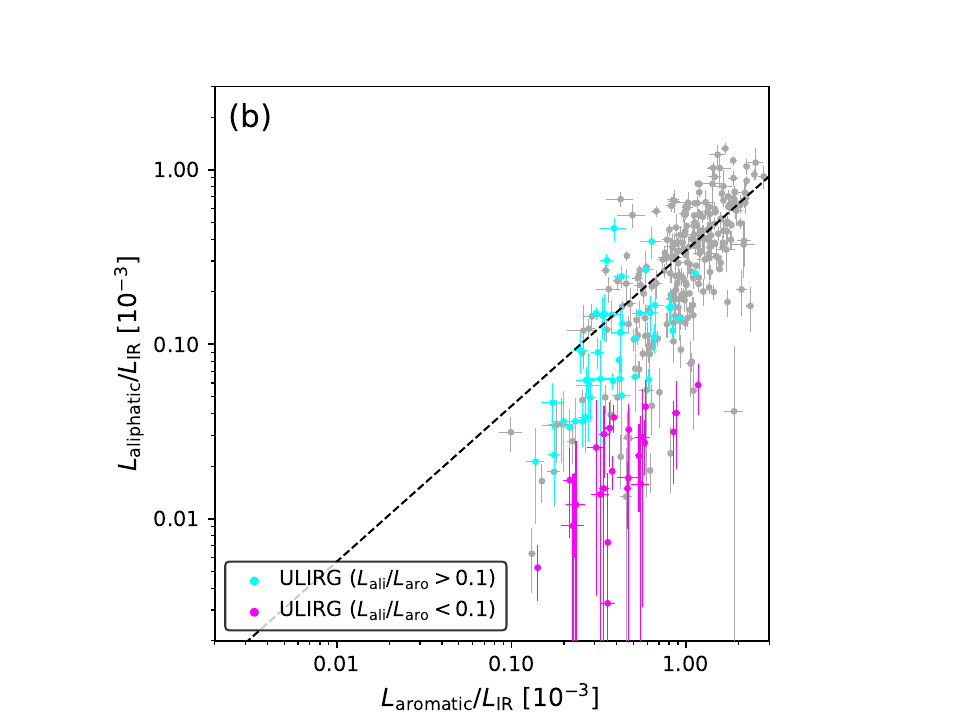}
\end{minipage} \\
\multicolumn{2}{c}{
    \begin{minipage}[c]{0.9\linewidth}
    \centering
    \includegraphics[bb=50 0 380 305, clip, width=0.45\linewidth]{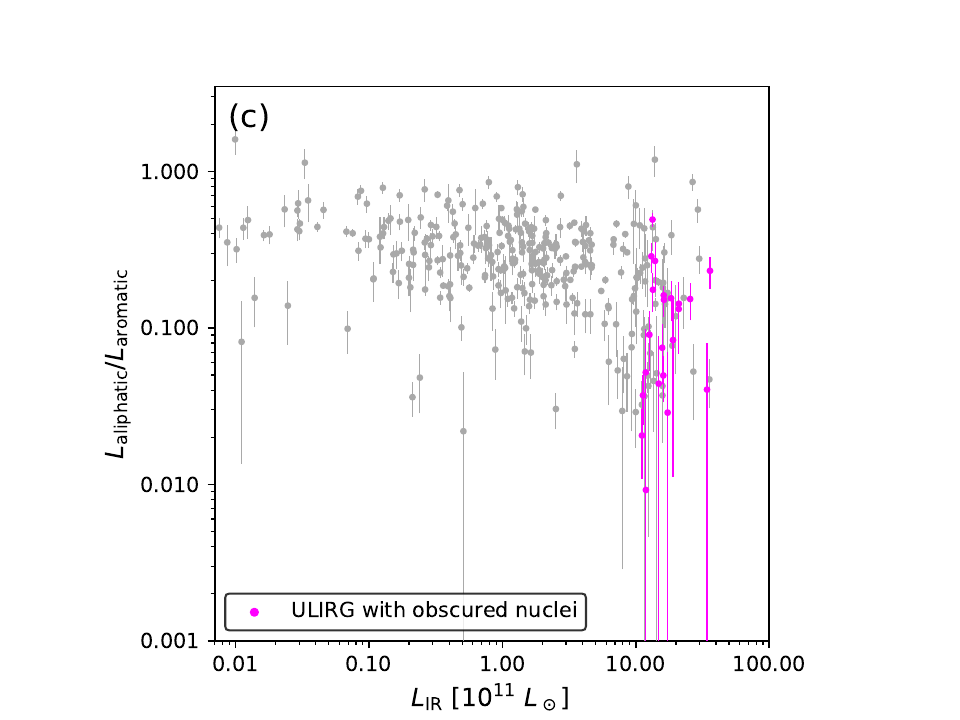}
    \end{minipage}
    }
\end{tabular}
\caption{(a) Same as figure \ref{aliaro_luminosityclass}c, but color-coded with $L_\mathrm{hot}/L_\mathrm{IR}$: $L_\mathrm{hot}/L_\mathrm{IR}>5.0\times10^{-3}$ (magenta circles) and $L_\mathrm{hot}/L_\mathrm{IR}<5.0\times10^{-3}$ (cyan circles). (b) Same as figure \ref{aliaro_lstarlir}b, but for ULIRG sample: $L_\mathrm{aliphatic}/L_\mathrm{aromatic}<$ 0.1 (magenta circles) and $L_\mathrm{aliphatic}/L_\mathrm{aromatic}>$ 0.1 (cyan circles). The (sub)IRG and LIRG samples are shown in grey color. The dashed line in panel(b) is the one fitted to the (sub-)IRG and LIRG data points. (c) Same as figure \ref{aliaro_5sigma}a, but ULIRGs with obscured nuclei based on the Spitzer/IRS mid-IR spectra (\cite{tsuchikawa2021}; \cite{donnan2023}) are denoted by magenta circles, while the others by grey circles. {Alt text: Three scatter plots.}}
\label{aliaro_ulirg}
\end{figure*}

Considering that the 4~\textmu m bluer continuum is likely to be related with old stellar populations, we utilize $L_\mathrm{star}$, which is derived from the SED fitting in section \ref{sed_fitting}, and examine the relationship between $F_\mathrm{4.2\ \mu m}/F_\mathrm{3.7\ \mu m}$ and $L_\mathrm{star}/L_\mathrm{IR}$. As shown in figure \ref{aliaro_lstarlir}a, galaxies with higher $L_\mathrm{star}/L_\mathrm{IR}$ show bluer continuum colors at 4~\textmu m, which indicates that the 4~\textmu m bluer continuum is indeed related to the stellar emission. \\
\indent The higher $L_\mathrm{aliphatic}/L_\mathrm{aromatic}$ seen for some of the (sub-)IRGs is caused by either higher $L_\mathrm{aliphatic}$ or lower $L_\mathrm{aromatic}$ (or both). Figure \ref{aliaro_lstarlir}b shows the relationship between $L_\mathrm{aliphatic}/L_\mathrm{IR}$ and $L_\mathrm{aromatic}/L_\mathrm{IR}$, where we can see that the distributions of the (sub-)IRG samples with $L_\mathrm{aliphatic}/L_\mathrm{aromatic}>$ 0.35 and $L_\mathrm{aliphatic}/L_\mathrm{aromatic}<$ 0.35 are separated along the vertical ($L_\mathrm{aliphatic}/L_\mathrm{IR}$) axis more clearly than along the horizontal ($L_\mathrm{aromatic}/L_\mathrm{IR}$) axis. Thus (sub-)IRGs with high $L_\mathrm{aliphatic}/L_\mathrm{aromatic}$ are likely to be caused by high $L_\mathrm{aliphatic}$, but not by low $L_\mathrm{aromatic}$. \\
\indent Then we revisit the result of figure \ref{aliaro_5sigma}b, $L_\mathrm{aliphatic}/L_\mathrm{aromatic}$ as a function of the UV radiation hardness. In figure \ref{aliaro_lstarlir}c, we plot only star-forming (sub-)IRGs together with the Galactic H\,\emissiontype{II} regions, where we can recognize a global decreasing trend more clearly. Hence, relatively high $L_\mathrm{aliphatic}$ is probably connected to the softer UV radiation, which is consistent with the large old stellar population (i.e., high $L_\mathrm{star}/L_\mathrm{IR}$). Since aliphatic hydrocarbons are more fragile than aromatic hydrocarbons for photodissociation, a relatively high abundance of aliphatic hydrocarbons is likely to reflect the intrinsic nature of PAHs outside the H\,\emissiontype{II} region where the PAHs are not irradiated by strong UV radiation fields and thus remain non-processed. In figure \ref{aliaro_lstarlir}d, we show the stacked spectra of the aromatic and aliphatic features of the (sub-)IRG sample, which may exhibit the authentic spectral features of UV-non-processed PAHs rich with aliphatic side groups as a form of the evolution of interstellar organic matter in galaxies.

\subsection{Possible causes for low $L_\mathrm{aliphatic}/L_\mathrm{aromatic}$ of ULIRGs}
\label{aliaro_high_luminosity}
As shown in figure \ref{aliaro_luminosityclass}c, the extremely low $L_\mathrm{aliphatic}/L_\mathrm{aromatic}$ for some ULIRGs is clearly related to the red continuum at 4~\textmu m and thus most probably the presence of a hot dust component. Then we utilize $L_\mathrm{hot}$, which is derived from the SED fitting, and examine the relationship between $F_\mathrm{4.2\ \mu m}/F_\mathrm{3.7\ \mu m}$ and $L_\mathrm{hot}/L_\mathrm{IR}$. As seen in figure \ref{aliaro_ulirg}a, galaxies with higher $L_\mathrm{hot}/L_\mathrm{IR}$ show redder continuum colors at 4~\textmu m, which indicates that the 4~\textmu m redder continuum is indeed related to the hot dust emission. A possibility of the origin of the hot dust component is AGNs, which escape our removal of AGN candidates from the ULIRG sample based on the near-IR spectra ($\mathrm{EW_{aromatic}}$ and $\Gamma$). For instance, galaxies with compact obscured nuclei (CONs; \cite{falstad2021}) can show continuum colors red at 4~\textmu m but relatively blue at 3~\textmu m (i.e., $\Gamma<$1). Another possibility is a galaxy merger, which can enhance starburst through the interaction of galaxies, producing hot dust components. In line with this, \citet{kondo2024} suggested that in addition to the ISRF strength, galaxy mergers with strong shocks may cause a further decrease in $L_\mathrm{aliphatic}/L_\mathrm{aromatic}.$ On the contrary, \citet{smith2007} pointed out that such hot dust as heated by the enhanced starburst cannot account for the continuum reddening at 4~\textmu m. Moreover, \citet{lyu2025} showed that galaxy mergers do not decrease $L_\mathrm{aliphatic}/L_\mathrm{aromatic}$ clearly using their JWST sample, although the luminosity classes of the galaxies are much different between our sample and their sample as shown in figure \ref{example_broad_feature}. \\
\indent In the same manner as in the previous subsection, the extremely low $L_\mathrm{aliphatic}/L_\mathrm{aromatic}$ ($<$ 0.1) seen for some of the ULIRGs is caused by either lower $L_\mathrm{aliphatic}$ or higher $L_\mathrm{aromatic}$ (or both). Figure \ref{aliaro_ulirg}b is the same as figure \ref{aliaro_lstarlir}b, but we focus on the ULIRG sample, where we can see that their distributions with $L_\mathrm{aliphatic}/L_\mathrm{aromatic}>$ 0.1 and $L_\mathrm{aliphatic}/L_\mathrm{aromatic}<$ 0.1 are separated along the vertical ($L_\mathrm{aliphatic}/L_\mathrm{IR}$) axis more clearly than along the horizontal ($L_\mathrm{aromatic}/L_\mathrm{IR}$) axis. Thus ULIRGs with extremely low $L_\mathrm{aliphatic}/L_\mathrm{aromatic}$ are likely to be caused by low $L_\mathrm{aliphatic}$, but not by high $L_\mathrm{aromatic}$. \\
\indent Past studies reported that some IR galaxies in merger systems show both starburst and AGN activities (e.g., Arp 299: \cite{ballo2004}; Mrk 273: \cite{nardini2008}, \yearcite{nardini2009}; NGC 3341: \cite{bianchi2013}; Mrk 266: \cite{beaulieu2023}; NGC 3256: \cite{bohn2024}; NGC 6240: \cite{hermosa2025}). The galactic nuclei in these merger systems are often heavily obscured, for which aliphatic hydrocarbon features are observed in the absorption at 3.4~\textmu m (e.g., \cite{imanishi2008}, \yearcite{imanishi2010}). Consequently, the observed $L_\mathrm{aliphatic}/L_\mathrm{IR}$ in ULIRGs that associate heavily obscured nuclei is likely to be underestimated, because the aliphatic emission features are blended with the aliphatic absorption features. Indeed, 33\% of the ULIRGs in our sample are likely to contain obscured nuclei, based on the Spitzer/IRS mid-IR spectra (\cite{tsuchikawa2021}; \cite{donnan2023}). Figure \ref{aliaro_ulirg}c shows the distribution of such ULIRGs in the $L_\mathrm{aliphatic}/L_\mathrm{aromatic}$ vs $L_\mathrm{IR}$ plane, from which we recognize that there is an indication of ULIRGs with obscured nuclei showing extremely low $L_\mathrm{aliphatic}/L_\mathrm{aromatic}$. To correctly estimate $L_\mathrm{aliphatic}/L_\mathrm{aromatic}$ in merger galaxies with heavily obscured nuclei, it is necessary to consider modeling both aliphatic emission and absorption features for each interacting galaxy with higher spatial and spectral resolutions spectra such as those taken with JWST/NIRSpec.

\section{Conclusions}
Using near-IR (2.55--4.85~\textmu m) spectra of nearby galaxies obtained with AKARI/IRC, we have systematically investigated the relationship between the luminosities of the aromatic hydrocarbon feature at 3.3~\textmu m ($L_\mathrm{aromatic}$) and the aliphatic hydrocarbon features at 3.4--3.6~\textmu m ($L_\mathrm{aliphatic}$) for 243 spectra of 240 star-forming (U)LIRGs, 119 spectra of 105 star-forming IRGs, and 94 spectra of 65 sub-IRGs, in addition to 232 spectra of 36 Galactic H\,\emissiontype{II} regions as a reference sample. As a result, $L_\mathrm{aliphatic}/L_\mathrm{aromatic}$ of the sample galaxies shows considerably large variations, compared to the Galactic H\,\emissiontype{II} regions, where $L_\mathrm{aliphatic}/L_\mathrm{aromatic}$ decreases systematically with the total infrared luminosity over the range of $10^9$--$10^{13}\ \mathrm{L_\odot}$. We find that (sub-)IRGs with continuum colors bluer at 4~\textmu m tend to have higher $L_\mathrm{aliphatic}/L_\mathrm{aromatic}$, while ULIRGs with continuum colors redder at 4~\textmu m tend to have extremely low $L_\mathrm{aliphatic}/L_\mathrm{aromatic}$. We show that the variations of $L_\mathrm{aliphatic}/L_\mathrm{aromatic}$ are mainly caused by the changes in $L_\mathrm{aliphatic}$ but not by those in $L_\mathrm{aromatic}$. It is probable that the (sub-)IRGs with high $L_\mathrm{aliphatic}/L_\mathrm{aromatic}$ are relatively inactive with star formation, having aliphatic-rich hydrocarbon dust, while the ULIRGs with extremely low $L_\mathrm{aliphatic}/L_\mathrm{aromatic}$ are merger systems with galactic nuclei, showing very low $L_\mathrm{aliphatic}$. We conclude that the former is likely to reflect the intrinsic nature of PAHs outside the H\,\emissiontype{II} region where the PAHs remain non-processed by strong UV radiation fields, while the latter is likely to be caused by blending aliphatic emission and absorption features due to the presence of an obscured galactic nucleus in merger systems.

\begin{ack}
This research is based on observations with AKARI, a JAXA project with the participation of ESA. This publication makes use of data products from the Wide-field Infrared Survey Explorer, which is a joint project of the University of California, Los Angeles, and the Jet Propulsion Laboratory/California Institute of Technology, funded by the National Aeronautics and Space Administration, and data products from the Infrared Astronomical Satellite (IRAS), which is a joint project of the US, UK, and the Netherlands. This research has made use of the NASA/IPAC Extragalactic Database, which is funded by the National Aeronautics and Space Administration and operated by the California Institute of Technology and use of the SIMBAD database, operated at CDS, Strasbourg, France. This work is financially supported by JST SPRING, Grant Number JPMJSP2125. The principal author is grateful for the “THERS Make New Standards Program for the Next Generation Researchers.” This work was supported by JSPS KAKENHI Grant Number 24K07094.
\end{ack}


\end{document}